\documentclass[
  aps,
  prd,
  11pt,
  superscriptaddress,
  notitlepage,
  onecolumn
]{revtex4-1}

\AtBeginDocument{
  \linespread{1.05}\selectfont
}

\usepackage[
  a4paper,
  left=2.3cm,
  right=2.3cm,
  top=2.5cm,
  bottom=2.5cm
]{geometry}

\usepackage{microtype} 
\usepackage{graphicx}
\usepackage{xcolor}
\usepackage{orcidlink}
\usepackage{enumerate}
\usepackage{enumitem}
\usepackage{ulem}
\usepackage{amssymb} 
\usepackage[svgnames]{xcolor} 
\usepackage{tikz}
\usepackage{mathtools}
\usepackage[dvipsnames]{xcolor}
\usetikzlibrary{arrows.meta,positioning,fit,calc}

\definecolor{linkcolor}{rgb}{0.0,0.3,0.5}
\definecolor{linkcolor}{rgb}{0.0,0.3,0.5}
\definecolor{mypurple}{RGB}{143, 116, 210}

\newcommand{\du}{\mathrm{d}}

\graphicspath{{Figures/}}

\newcommand{\hias}{School of Fundamental Physics and Mathematical Sciences, Hangzhou Institute for Advanced Study, University of Chinese Academy of Sciences, Hangzhou 310024, China}

\usepackage{ytableau}

\usepackage{soul}

\begin{document}

\title{Neural variational framework for random Young-diagram limit shapes}

\author{Qian Chen
\orcidlink{0000-0002-7705-9698}}
\email{chenqian.phys@gmail.com}
\affiliation{Mathematics Research Center, School of Science and Engineering, The Chinese University of Hong Kong, Shenzhen, China}

\author{Bo-Xuan Ge
\orcidlink{0000-0003-0738-3473}}
\email{bo-xuan.ge@ucas.ac.cn}
\affiliation{\hias}

\begin{abstract}
We develop a structure-preserving neural variational framework for
random Young-diagram ensembles, with representations adapted to the
structure and scaling of each measure.  The method is validated on the
Plancherel, uniform, minimal-difference, and fixed-\(q\)
\(q\)-Plancherel ensembles, using known asymptotic profiles only for
post-training comparison.  We then study a quartically deformed
hook-length ensemble without assuming an analytical saddle shape.
Large-\(n\) neural profiles are compared with finite-size MAP
profiles obtained from exact-action searches and with mean profiles
obtained from corner-transfer Metropolis--Hastings sampling.
Increasing the deformation suppresses
the leading rows and broadens the support, while the neural, discrete,
and sampled mean profiles agree at the percent level.  These results
provide numerical evidence for a deformation-dependent macroscopic
saddle family.
\end{abstract}

\maketitle

\section{Introduction}
\label{sec:introduction}

For each \(n\in\mathbb{N}\), the irreducible complex representations of
the symmetric group \(S_n\) are indexed by partitions
\(\lambda\vdash n\), or equivalently by Young diagrams with \(n\)
boxes~\cite{fulton1991representation}.  Young diagrams also appear
throughout representation theory, probability, mathematical physics,
and quantum information.  A familiar example is Schur--Weyl duality,
under which
\((\mathbb{C}^{d})^{\otimes n}\) decomposes into irreducible components
of \(S_n\) and \(\mathrm{GL}(d,\mathbb{C})\), labeled by partitions with
at most \(d\) rows~\cite{bacon2005Schur}.

The dimension of the irreducible representation indexed by
\(\lambda\) is determined by the hook lengths of the diagram through
the hook-length formula.  The regular representation of \(S_n\) then
induces the Plancherel measure
\(\mathbb{P}^{\rm Pl}_n(\lambda)=(\dim\lambda)^2/n!\).
More generally, other choices of weights define different probability
measures on the set of partitions of \(n\), and hence different random
Young-diagram ensembles.

For a sequence of such measures, a central asymptotic question is
whether a suitably rescaled random diagram concentrates around a
deterministic profile as \(n\to\infty\).  Two objects should be
distinguished.  At fixed \(n\), we call any maximizer
\[
    \lambda_{\rm MAP}(n)
    \in
    \operatorname*{arg\,max}_{\lambda\vdash n}
    \mathbb{P}_n(\lambda)
\]
a maximum-probability diagram.  We use MAP, short for
maximum a posteriori, as a convenient name for this object, although no
Bayesian inference is involved here.  A typical limit shape, by
contrast, describes the macroscopic region in profile space where the
total probability mass concentrates.  The two notions need not agree
for a general measure.

For the Plancherel measure, maximizing the probability is equivalent to
maximizing \(\dim\lambda\).  The classical works of Logan and Shepp and
of Vershik and Kerov showed that the extremal and typical problems are
governed by the same VKLS macroscopic profile
\cite{logan1977variational,verKer77ap,vershik1985lt}.
Related concentration phenomena occur for many other partition
ensembles, although both the limiting profile and the appropriate
scaling may depend on the measure~\cite{mkrtchyan2012mt}.

A range of analytical methods has been developed for these problems.
The classical limit-shape results rely on variational arguments, while
the Plancherel measure also admits a determinantal treatment based on
poissonization, depoissonization, and asymptotic analysis of the
discrete Bessel kernel~\cite{borodin2000apm}.  Such methods are
particularly effective when the measure possesses sufficient algebraic
or integrable structure.  For deformed hook-weighted ensembles,
however, the effective action can remain nonlocal and an analytical
saddle profile may not be available.  Direct optimization over integer
partitions is still possible at finite \(n\), but becomes increasingly
costly when one seeks a smooth profile at much larger sizes.
Neural-network parameterizations have also been used as direct
numerical ans\"atze for variational problems, notably in the Deep Ritz
method~\cite{e2018deep}.

In this work, we develop a structure-preserving neural variational
framework for random Young-diagram ensembles.  Rather than imposing one
common representation on all measures, we choose the variational
variables according to the structure and scaling of the ensemble.
Finite-size hook actions are used for hook-weighted measures,
coarse-grained densities for entropy-dominated ensembles, and discrete
row fractions when the leading rows scale linearly with \(n\).
Monotonicity, fixed area, and the reconstruction of an integer partition
are incorporated into the parametrization or enforced through
projection.

We first test the framework on ensembles with known analytical or
asymptotic behavior: the Plancherel measure, uniform random partitions,
minimal-difference partitions, and the fixed-\(q\)
\(q\)-Plancherel ensemble.  The known profiles are used only for
post-training validation and do not enter the optimization objectives or
checkpoint selection.  These examples test both the accuracy of the
neural approximation and its ability to adapt to different asymptotic
scalings.

We then apply the same strategy to a quartically deformed hook-length
ensemble for which no closed-form saddle profile is assumed.  
The large-\(n\) neural profiles are compared with finite-size MAP
profiles obtained from exact-action searches over integer partitions
and with mean profiles obtained from corner-transfer
Metropolis--Hastings sampling of the finite-temperature ensemble.
The
three calculations produce closely agreeing deformation-dependent
profiles, providing independent numerical evidence for the corresponding
macroscopic saddle family.





\section{Young-diagram ensembles and variational limit shapes}
\label{sec:ensembles}

\subsection{Partitions and Young-diagram profiles}

A partition of a positive integer \(n\) is a finite nonincreasing
sequence of positive integers,
\begin{equation}
    \lambda=(\lambda_1,\lambda_2,\ldots,\lambda_{\ell(\lambda)}),
    \qquad
    \lambda_1\geq\lambda_2\geq\cdots
    \geq\lambda_{\ell(\lambda)}>0,
    \qquad
    |\lambda|=\sum_{i=1}^{\ell(\lambda)}\lambda_i=n .
\end{equation}
Here \(\ell(\lambda)\) is the number of nonzero parts.  When convenient,
we set \(\lambda_i=0\) for \(i>\ell(\lambda)\), but omit these trailing
zeros from the notation.  We write \(\lambda\vdash n\), and denote the
set of partitions of \(n\) by \(\mathbb{Y}_n\).

We use the English convention for Young diagrams.  The coordinate
\(u=(i,j)\) labels the box in the \(i\)th row and \(j\)th column, with
rows counted from top to bottom and columns from left to right.  Thus
\(u=(i,j)\) belongs to \(\lambda\) when
\begin{equation}
    1\leq j\leq\lambda_i .
\end{equation}
For example, the partition \((3,2)\) is represented by
\begin{align*}
\begin{ytableau}
    \ & \ & \ \\
    \ & \
\end{ytableau}.
\end{align*}
In the profile convention used below, the positive \(x\)-direction runs
downward along the rows.

For balanced Young diagrams, both the largest row and the number of
nonzero rows are of order \(\sqrt n\).  We use the scaled English
coordinates
\begin{equation}
    x_i=\frac{i}{\sqrt n},
    \qquad
    f_{\lambda,n}(x_i)=\frac{\lambda_i}{\sqrt n}.
    \label{eq:english_profile}
\end{equation}
The finite-\(n\) profile is understood as the step function
\begin{equation}
    f_{\lambda,n}(x)
    =
    \frac{\lambda_i}{\sqrt n},
    \qquad
    \frac{i-1}{\sqrt n}<x\leq\frac{i}{\sqrt n}.
\end{equation}
It is nonnegative and nonincreasing, and its area satisfies
\begin{equation}
    \int_0^\infty f_{\lambda,n}(x)\,\mathrm{d}x
    =
    \frac{1}{n}\sum_{i\geq1}\lambda_i
    =
    1 .
\end{equation}

Other ensembles may require a different macroscopic scaling.  We therefore
also allow
\begin{equation}
    x_i=\frac{i}{a_n},
    \qquad
    f_{\lambda,n}(x)
    =
    \frac{\lambda_i}{b_n},
    \qquad
    \frac{i-1}{a_n}<x\leq\frac{i}{a_n},
    \qquad
    a_n b_n=n .
    \label{eq:anisotropic_scaling}
\end{equation}
The area normalization remains
\begin{equation}
    \int_0^\infty f_{\lambda,n}(x)\,\mathrm{d}x
    =
    \frac{1}{a_n}
    \sum_{i\geq1}\frac{\lambda_i}{b_n}
    =
    1 .
\end{equation}
For the fixed-\(q\) \(q\)-Plancherel ensemble with \(0<q<1\), the
leading rows can instead be of order \(n\).  In that case it is more
natural to keep the row index \(i\) discrete and work with the row
fractions
\begin{equation}
    \mu_i^{(n,q)}=\frac{\lambda_i}{n},
    \qquad
    \sum_{i\geq1}\mu_i^{(n,q)}=1 .
\end{equation}

\subsection{Probability measures on Young diagrams}

An $n$-box Young-diagram ensemble is a family of probability measures
\begin{equation}
    \mathbb{P}_{n,\theta}(\lambda),
    \qquad
    \lambda\in\mathbb{Y}_n,
    \label{eq:general_measure}
\end{equation}
where \(\theta\) collects the parameters of the ensemble.  We are
interested in the large-\(n\) behavior of the rescaled profile
\(f_{\lambda,n}\) when \(\lambda\) is sampled from
\(\mathbb{P}_{n,\theta}\).  A deterministic profile \(f_\theta\) is
called the limit shape if
\begin{equation}
    f_{\lambda,n}\longrightarrow f_\theta
    \qquad
    \text{in probability as } n\to\infty .
\end{equation}
For precise definitions of Young-diagram limit shapes, see
Refs.~\cite{logan1977variational,verKer77ap,vershik1985lt,mkrtchyan2012mt}.


The ensembles considered in this paper fall into several classes.

\subsubsection{Plancherel measure}
The Plancherel measure is defined by
\begin{equation}
    \mathbb{P}^{\rm Pl}_n(\lambda)
    =
    \frac{(\dim\lambda)^2}{n!},
    \qquad
    \lambda\vdash n,
    \label{eq:plancherel_measure}
\end{equation}
where \(\dim\lambda\) is the dimension of the irreducible representation
of \(S_n\) indexed by \(\lambda\).  The identity
\begin{equation}
    \sum_{\lambda\vdash n}(\dim\lambda)^2=n!
\end{equation}
ensures that the measure is normalized.

By the hook-length formula~\cite{fulton1991representation},
\begin{equation}
    \dim\lambda
    =
    \frac{n!}{\displaystyle\prod_{(i,j)\in\lambda}h(i,j)},
    \qquad
    h(i,j)
    =
    \lambda_i-j+\lambda'_j-i+1,
    \label{eq:hook_length_formula}
\end{equation}
where \(\lambda'\) is the conjugate partition obtained by transposing the
Young diagram.  For example,
\begin{align*}
\left(
\begin{ytableau}
    \ & \ & \ \\
    \ & \
\end{ytableau}
\right)'
=
\begin{ytableau}
    \ & \ \\
    \ & \ \\
    \
\end{ytableau}.
\end{align*}

Since \(h(i,j)\) depends on both the row length \(\lambda_i\) and the
column length \(\lambda'_j\), the hook contribution couples the row and
column structures of the diagram. Substituting
Eq.~\eqref{eq:hook_length_formula} into
Eq.~\eqref{eq:plancherel_measure} and taking the negative logarithm
gives
\begin{equation}
    -\log\mathbb{P}^{\rm Pl}_n(\lambda)
    =
    2\sum_{(i,j)\in\lambda}\log h(i,j)
    -
    \log n! .
\end{equation}
Dropping the additive term \(-\log n!\), we define the finite-size
Plancherel action by
\begin{equation}
    H^{\rm Pl}_n(\lambda)
    =
    2\sum_{(i,j)\in\lambda}\log h(i,j).
    \label{eq:plancherel_action}
\end{equation}


\subsubsection{Uniform random partitions}
The uniform ensemble assigns equal probability to every partition of
\(n\):
\begin{equation}
    \mathbb{P}^{\rm unif}_n(\lambda)
    =
    \frac{1}{p(n)},
    \qquad
    \lambda\vdash n,
    \label{eq:uniform_measure}
\end{equation}
where \(p(n)=|\mathbb{Y}_n|\) denotes the partition number.
No closed-form expression is known, although asymptotic formulas such
as the Hardy--Ramanujan formula are available.
Since all partitions have the same finite-size weight, there is no distinguished
MAP partition.  Nevertheless, the ensemble has a nontrivial typical
limit shape~\cite{vershik1996statistical}.  This shape is selected by the entropy of the microscopic
partitions associated with a given macroscopic profile, rather than by
variations in their individual probability weights.


\subsubsection{Minimal-difference-\(p\) partitions}
\label{subsubsec:minimal_difference_p}

For \(p\in\mathbb{Z}_{\geq 0}\), define
\begin{equation}
    \mathbb{Y}^{(p)}_n
    =
    \left\{
    \lambda\vdash n:
    \lambda_i-\lambda_{i+1}\geq p,
    \quad
    i=1,\ldots,\ell(\lambda)-1
    \right\}.
    \label{eq:minimal_difference_set}
\end{equation}
The corresponding ensemble is uniform on this constrained set:
\begin{equation}
    \mathbb{P}^{(p)}_n(\lambda)
    =
    \frac{1}{Z_{n,p}}
    \mathbf{1}_{\{\lambda\in\mathbb{Y}^{(p)}_n\}},
    \qquad
    Z_{n,p}=|\mathbb{Y}^{(p)}_n|,
    \label{eq:minimal_difference_measure}
\end{equation}
where \(\mathbf{1}_{A}\) denotes the indicator of the set \(A\).
The cases \(p=0\) and \(p=1\) give ordinary and distinct partitions,
respectively.  As \(p\) increases, consecutive parts are forced farther
apart, giving a \(p\)-dependent family of constrained partition ensembles
with an exclusion-statistics interpretation~\cite{comtet2007integer,comtet2008note}.


\subsubsection{\(q\)-Plancherel measure}
\label{subsubsec:q_plancherel_measure}

We use the \(q\)-Plancherel ensemble as a test of whether the relevant
large-\(n\) scaling can change with the measure.  It is the
Hecke-algebra deformation of the ordinary Plancherel measure~\cite{feray2012asymptotics}.  For
\(0<q<1\), define
\begin{equation}
    [m]_q=\frac{1-q^m}{1-q},
    \qquad
    [n]_q! = \prod_{m=1}^n [m]_q ,
    \label{eq:q_integer_factorial}
\end{equation}
with \([m]_1=m\) understood by continuity.  For
\(\lambda\vdash n\), let
\begin{equation}
    b(\lambda)
    =
    \sum_{i\geq1}(i-1)\lambda_i .
    \label{eq:b_lambda_definition}
\end{equation}
The generic degree associated with \(\lambda\) is
\begin{equation}
    D_\lambda(q)
    =
    \frac{
        q^{b(\lambda)}[n]_q!
    }{
        \displaystyle\prod_{u\in\lambda}[h(u)]_q
    },
    \label{eq:generic_degree}
\end{equation}
and the \(q\)-Plancherel measure is
\begin{equation}
    \mathbb{P}^{q{\rm Pl}}_{n,q}(\lambda)
    =
    \frac{D_\lambda(q)\,\dim\lambda}{[n]_q!},
    \qquad
    \lambda\vdash n .
    \label{eq:q_plancherel_measure_definition}
\end{equation}
Using the ordinary hook-length formula, this becomes
\begin{equation}
    \mathbb{P}^{q{\rm Pl}}_{n,q}(\lambda)
    =
    n!\,
    q^{b(\lambda)}
    \prod_{u\in\lambda}
    \frac{1}{h(u)[h(u)]_q}.
    \label{eq:q_plancherel_measure_hook_form}
\end{equation}
After dropping terms independent of \(\lambda\), the corresponding
finite-size action is
\begin{equation}
    H^{q{\rm Pl}}_n(\lambda)
    =
    \sum_{u\in\lambda}\log h(u)
    +
    \sum_{u\in\lambda}\log [h(u)]_q
    -
    \log q\, b(\lambda).
    \label{eq:q_plancherel_action}
\end{equation}
The last term is part of the measure itself.  Since \(\log q<0\) for
\(0<q<1\), it penalizes boxes in lower rows and biases the diagram toward
its first few rows.

The ordinary Plancherel measure is recovered as \(q\to1\).  Indeed,
\([h(u)]_q\to h(u)\) and \(q^{b(\lambda)}\to1\), so that
\begin{equation}
    \mathbb{P}^{q{\rm Pl}}_{n,q}(\lambda)
    \longrightarrow
    n!\prod_{u\in\lambda}\frac{1}{h(u)^2}
    =
    \mathbb{P}^{\rm Pl}_n(\lambda).
\end{equation}

For fixed \(0<q<1\), the large-\(n\) scaling differs from the balanced
Plancherel regime.  The leading rows are of order \(n\), and for each
fixed \(i\),
\begin{equation}
    \frac{\lambda_i}{n}
    \longrightarrow
    (1-q)q^{i-1}
    \qquad
    \text{in probability}
    \label{eq:q_plancherel_row_fractions}
\end{equation}
as \(n\to\infty\)~\cite{feray2012asymptotics}.
The natural macroscopic variables are therefore the discrete row
fractions \(\lambda_i/n\), rather than a continuous profile obtained
from the usual \(\sqrt n\) scaling.


\subsubsection{Quartically deformed hook-length ensemble}
\label{subsubsec:quartic_hook_ensemble}

We finally introduce a deformed ensemble for which no analytic limit
shape is assumed.  It serves as the main non-benchmark application of
the neural variational solver developed in this work.

For \(\theta\geq0\) and \(c_h>0\), we define the finite-size action
\begin{equation}
    S^{(c_h)}_{n,\theta}(\lambda)
    \coloneqq
    c_h
    \sum_{u\in\lambda}\log h(u)
    +
    \theta\sqrt n
    \sum_{i\geq1}
    \left(
        \frac{\lambda_i}{\sqrt n}
    \right)^4,
    \qquad
    \lambda\vdash n .
    \label{eq:quartic_hook_action}
\end{equation}
The corresponding probability measure is
\begin{equation}
    \mathbb{P}^{(c_h,\theta)}_n(\lambda)
    =
    \frac{1}{Z_{n,\theta}^{(c_h)}}
    \exp\!\left[
        -S^{(c_h)}_{n,\theta}(\lambda)
    \right],
    \qquad
    Z_{n,\theta}^{(c_h)}
    =
    \sum_{\lambda\vdash n}
    \exp\!\left[
        -S^{(c_h)}_{n,\theta}(\lambda)
    \right].
    \label{eq:quartic_hook_measure}
\end{equation}
The first term is the hook-length contribution, while the second is a
quartic potential acting on the row lengths in the balanced scaling.

The prefactor \(\sqrt n\) is chosen so that the deformation contributes
at the same variational order as the shape-dependent part of the hook
action.  For a balanced diagram,
\begin{equation}
    \frac{\lambda_i}{\sqrt n}=O(1),
    \qquad
    \ell(\lambda)=O(\sqrt n),
\end{equation}
and hence
\begin{equation}
    \sum_{i\geq1}
    \left(
        \frac{\lambda_i}{\sqrt n}
    \right)^4
    =
    O(\sqrt n).
\end{equation}
The quartic term is therefore \(O(n)\) after multiplication by
\(\sqrt n\).  An \(O(1)\) value of \(\theta\) can consequently modify the
large-\(n\) saddle shape.

The coefficient \(c_h\) fixes the normalization of the hook term.  Since
\begin{equation}
    -\log\mathbb{P}^{\rm Pl}_n(\lambda)
    =
    2\sum_{u\in\lambda}\log h(u)
    -\log n!,
\end{equation}
the Plancherel normalization corresponds to \(c_h=2\).  The numerical
results in this work use \(c_h=1\).  At \(\theta=0\), this choice reduces
to the Gelfand measure, whose weights are proportional to
\(\dim\lambda\) and whose first-order limit shape is again the VKLS
profile~\cite{meliot2010gelfand}.  We therefore refer to the deformed
model as a quartically deformed hook-length ensemble rather than as a
strict deformation of the Plancherel measure.

The quartic term penalizes configurations containing very long rows.
Increasing \(\theta\) is therefore expected to suppress
\(\lambda_1/\sqrt n\) and redistribute boxes over a broader range of
rows.  

\subsection{MAP diagrams and typical limit shapes}
\label{subsec:maximum_probability_typical}

For fixed \(n\), a maximum-a-posteriori (MAP) diagram is defined by
\begin{equation}
    \lambda_{\mathrm{MAP}}(n,\theta)
    \in
    \operatorname*{arg\,max}_{\lambda\in\mathbb{Y}_n}
    \mathbb{P}_{n,\theta}(\lambda).
    \label{eq:maximum_probability_diagram}
\end{equation}
The maximizer need not be unique.  A MAP diagram concerns the probability
of a single finite partition, whereas a limit shape describes the
concentration of the full measure on rescaled profiles.  The two may have
the same leading large-\(n\) profile, but they are not equivalent
definitions.

For the Plancherel measure, maximizing the probability is equivalent to
minimizing the hook-length action.  Indeed,
\begin{equation}
    -\log\mathbb{P}^{\rm Pl}_n(\lambda)
    =
    H^{\rm Pl}_n(\lambda)-\log n!,
\end{equation}
where \(H^{\rm Pl}_n\) is defined in
Eq.~\eqref{eq:plancherel_action}.  In this case the measure becomes
strongly concentrated as \(n\) grows, and the MAP sequence and typical
random diagrams approach the same leading limit shape.

The uniform ensemble shows why this relation cannot be assumed in
general.  Since every partition has the same probability,
\begin{equation}
    \operatorname*{arg\,max}_{\lambda\in\mathbb{Y}_n}
    \mathbb{P}^{\rm unif}_n(\lambda)
    =
    \mathbb{Y}_n .
\end{equation}
The MAP problem is therefore completely degenerate, although a typical
random partition still has a nontrivial limit shape.  The latter is
selected by the number of microscopic partitions lying near a given
macroscopic profile, rather than by differences between their individual
weights.  The same distinction applies to the uniform measure on
minimal-difference-\(p\) partitions.

We therefore keep the terms MAP diagram and typical limit shape distinct
throughout this work.  For hook-weighted ensembles, finite-\(n\) action
minimization provides a candidate saddle profile.  In the Plancherel case,
its leading limit is known to agree with the typical shape; for the
quartically deformed ensemble, this relation is tested numerically by
comparing the MAP profile with the MCMC mean.  For uniform and
minimal-difference ensembles, the relevant variational problem instead
includes the entropy of the ensemble.

\subsection{Variational formulation of limit shapes}
\label{subsec:variational_limit_shapes}

The large-\(n\) limit shape can often be characterized through a
variational principle.  Suppose that the induced distribution of
rescaled profiles satisfies a large-deviation principle of the form
\begin{equation}
    \mathbb{P}_{n,\theta}
    \left(
        f_{\lambda,n}\approx f
    \right)
    \asymp
    \exp\left[
        -A_n\mathcal{I}_\theta[f]
    \right],
    \label{eq:ldp_form}
\end{equation}
where \(A_n\) is the large-deviation speed and
\(\mathcal{I}_\theta[f]\) is the corresponding rate functional.  A limit
shape is then obtained from
\begin{equation}
    f_\theta
    \in
    \operatorname*{arg\,min}_{f\in\mathcal{A}}
    \mathcal{I}_\theta[f].
    \label{eq:limit_shape_variational}
\end{equation}
When the minimizer is unique, the measure concentrates on the profile
\(f_\theta\).

In the balanced scaling, the admissible profiles satisfy
\begin{equation}
    \mathcal{A}
    =
    \left\{
        f:[0,\infty)\to[0,\infty):
        f \text{ is nonincreasing},
        \quad
        \int_0^\infty f(x)\,\mathrm{d}x=1
    \right\}.
    \label{eq:admissible_set}
\end{equation}
This form of the admissible set is specific to the balanced continuum
scaling.  Ensembles such as fixed-\(q\) \(q\)-Plancherel require a
different macroscopic description.

Depending on the ensemble, the rate functional may contain contributions
associated with both the finite-size weight and the number of microscopic
partitions realizing a given macroscopic profile.  Schematically, one may
write
\begin{equation}
    \mathcal{I}_\theta[f]
    =
    \mathcal{E}_\theta[f]
    -
    \mathcal{S}_\theta[f],
    \label{eq:free_energy_schematic}
\end{equation}
with any ensemble-dependent coefficients absorbed into
\(\mathcal{E}_\theta\) and \(\mathcal{S}_\theta\).  For hook-weighted
ensembles, the finite-size action supplies the principal variational
contribution.  For uniform and minimal-difference partitions, the
macroscopic profile is instead selected through the entropy of the
ensemble.

The neural method developed in the next section follows this formulation.
The network provides a structure-preserving parameterization of the
admissible profile, while the objective is constructed from the defining
finite-size action or continuum entropy functional.  Analytical limit
shapes, when available, are used only for validation.

\section{Structure-preserving variational framework and
discrete validation}
\label{sec:method}

All five ensembles introduced in Sec.~\ref{sec:ensembles} are treated
with structure-preserving neural solvers.  Their numerical
representations and objectives, however, depend on the statistical
structure of the ensemble.  Plancherel and the quartically deformed
hook-length ensemble are formulated through differentiable finite-size
hook actions.  Uniform partitions use a density-first formulation based
on the Bose-type entropy, while minimal-difference-\(p\) partitions are
treated with an inverse Euler--Lagrange neural solver.  For fixed-\(q\)
\(q\)-Plancherel, the network acts directly on a finite set of row
fractions, rather than on a continuum profile in the balanced
\(\sqrt n\) scaling.

In each case, the Young-diagram constraints are built into the neural
representation.  The row-based solvers produce nonnegative,
nonincreasing row lengths with the required normalization.  In the
density-based calculations, the network produces a nonnegative density,
from which the corresponding nonincreasing profile is reconstructed.
Analytical limit shapes do not enter the training objectives or
checkpoint selection.  When available, they are used only in
post-optimization comparisons.

No analytical profile is assumed for the quartically deformed
hook-length ensemble.  The neural calculation provides a large-\(n\)
relaxed profile, which is supplemented by a finite-size MAP search based
on the exact integer action and by fixed-temperature corner-transfer
Metropolis--Hastings sampling.  The MAP calculation probes the
integer-partition saddle, while the MCMC calculation tests whether the
sampled mean remains close to the same profile.  The comparison of the
three calculations is presented in Sec.~\ref{sec:quartic_results}.
Further implementation details are given in
Appendix~\ref{app:method_details}.

\subsection{Structure-preserving neural representations}
\label{subsec:monotone_neural_ansatz}

For the balanced hook-action calculations, the network represents a
nonnegative generator,
\begin{equation}
    q_\phi(x)
    =
    {\rm softplus}\!\left(N_\phi(x)\right)
    +
    \epsilon_q,
    \qquad
    \epsilon_q>0.
    \label{eq:method_q_phi}
\end{equation}
The profile is constructed from its tail,
\begin{equation}
    f_\phi(x)
    =
    C_\phi
    \int_x^{X_{\max}}
    q_\phi(s)\,\du s,
    \label{eq:method_monotone_ansatz}
\end{equation}
where \(C_\phi\) fixes the area.  This representation is nonnegative and
nonincreasing by construction, with
\begin{equation}
    \int_0^{X_{\max}}f_\phi(x)\,\du x=1.
\end{equation}

The finite-grid realization depends on the ensemble.  The Plancherel and
quartic solvers act on continuous row lengths, while fixed-\(q\)
\(q\)-Plancherel is represented by a finite set of row fractions.  The
entropy-based calculations instead work with nonnegative densities, from
which the Young-diagram boundary is reconstructed.

A parameter-conditioned network is used for the
minimal-difference-\(p\) family.  Separate networks are trained for the
quartic deformation strengths, with continuation in \(\theta\) transferring
only the network parameters.  MAP and MCMC data are not used during neural
training.  The corresponding finite-grid definitions are given in
Appendix~\ref{app:hook_neural_details}.

\subsection{Differentiable hook-action objectives}
\label{subsec:method_soft_hook}

For continuous row lengths, the discrete box occupancy is replaced by
\begin{equation}
    O_{ij}^{(\phi)}
    =
    \sigma\left(
        \frac{
            \lambda_i^{(\phi)}
            -j+\delta_{\rm occ}
        }{\tau}
    \right),
    \qquad
    \sigma(z)=\frac{1}{1+e^{-z}},
    \label{eq:method_soft_occupancy}
\end{equation}
where \(\tau>0\) controls the width of the softened boundary and
\(\delta_{\rm occ}\) specifies the grid convention.  The corresponding
column lengths and hook lengths are
\begin{equation}
    \lambda_j^{\prime(\phi)}
    =
    \sum_i O_{ij}^{(\phi)},
    \label{eq:method_soft_columns}
\end{equation}
and
\begin{equation}
    h_{ij}^{(\phi)}
    =
    {\rm softplus}
    \left(
        \lambda_i^{(\phi)}
        -j
        +
        \lambda_j^{\prime(\phi)}
        -i
        +1
    \right)
    +
    \epsilon_h .
    \label{eq:method_soft_hook}
\end{equation}
As \(\tau\) is lowered, the occupancy approaches that of a discrete
Young diagram.

It is useful to denote the resulting soft hook contribution by
\begin{equation}
    \mathcal{H}^{(n)}_{\rm soft}(\phi)
    =
    \frac{1}{n}
    \sum_{i,j}
    O_{ij}^{(\phi)}
    \log h_{ij}^{(\phi)} .
    \label{eq:method_common_soft_hook}
\end{equation}
The ordinary Plancherel calculation minimizes
\begin{equation}
    \mathcal{L}^{(n)}_{\rm Pl}(\phi)
    =
    \mathcal{H}^{(n)}_{\rm soft}(\phi)
    +
    \mathcal{R}_{\rm Pl}.
    \label{eq:method_plancherel_soft_loss}
\end{equation}
The overall factor of two is omitted from the hook term because it does
not affect the minimizer of the unregularized Plancherel action.  The
weak numerical correction \(\mathcal{R}_{\rm Pl}\) is defined relative
to this normalization.

For \(q\)-Plancherel, the \(q\)-hook contribution and the row-dependent
term inherited from the exact measure are retained:
\begin{equation}
    \mathcal{L}^{(n)}_{q{\rm Pl}}(\phi;q)
    =
    {}
    \mathcal{H}^{(n)}_{\rm soft}(\phi)
    +
    \frac{1}{n}
    \sum_{i,j}
    O_{ij}^{(\phi)}
    \log [h_{ij}^{(\phi)}]_q    
    -
    \frac{\log q}{n}
    \sum_i(i-1)\lambda_i^{(\phi)}
    +
    \mathcal{R}_{q{\rm Pl}} .
    \label{eq:method_qplancherel_soft_loss}
\end{equation}

The quartically deformed calculation uses
\begin{equation}
\begin{split}
    \mathcal{L}^{(n)}_{\rm quartic}(\phi;\theta)
    =
    {}&
    c_h\,
    \mathcal{H}^{(n)}_{\rm soft}(\phi)
    +
    \frac{\theta}{\sqrt n}
    \sum_i
    \left(
        \frac{\lambda_i^{(\phi)}}{\sqrt n}
    \right)^4
    +
    \mathcal{R}_{\rm quartic}.
\end{split}
    \label{eq:method_quartic_neural_loss}
\end{equation}
The reported quartic calculations use \(c_h=1\), matching the discrete
ensemble defined in Sec.~\ref{subsubsec:quartic_hook_ensemble}.

The terms denoted by \(\mathcal{R}\) are weak numerical corrections
associated with the softened occupancy and finite computational domain.
They contain no analytical profile, MAP partition, or MCMC data.  After
optimization, the continuous rows are projected to an integer partition
and evaluated with the corresponding exact discrete action.  The grid
conventions, regularization terms, and temperature schedules are given in
Appendix~\ref{app:hook_neural_details}.

\subsection{Entropy-based objectives}
\label{subsec:method_entropy}

For uniformly weighted ensembles, the probability of an individual
partition does not select the typical profile.  We instead work with the
nonnegative density
\begin{equation}
    \rho(x)=-f'(x),
\end{equation}
for which
\begin{equation}
    f(x)=\int_x^\infty \rho(s)\,\du s,
    \qquad
    \int_0^\infty x\rho(x)\,\du x=1.
    \label{eq:method_density_area}
\end{equation}
On the finite grid, the profile is reconstructed from the corresponding
tail sum.

For ordinary uniform partitions, the relevant Bose-type entropy density
is
\begin{equation}
    s_{\rm B}(\rho)
    =
    (1+\rho)\log(1+\rho)
    -
    \rho\log\rho .
    \label{eq:method_bose_entropy_density}
\end{equation}
The neural density is normalized by the discrete area condition and
optimized using
\begin{equation}
    \mathcal{L}_{\rm unif}(\phi)
    =
    -
    \Delta x
    \sum_k
    s_{\rm B}\!\left(\rho_{\phi,k}\right)
    +
    \mathcal{R}_{\rm unif},
    \label{eq:method_uniform_entropy_loss}
\end{equation}
where \(\mathcal{R}_{\rm unif}\) contains only weak finite-grid
stabilizers.

Minimal-difference-\(p\) partitions are governed by the
exclusion-statistics entropy density~\cite{comtet2007integer,comtet2008note}
\begin{equation}
    s_p(\rho)
    =
    {}
    \left[1+(1-p)\rho\right]
    \log\left[1+(1-p)\rho\right]
    -
    \rho\log\rho
    -
    \left[1-p\rho\right]
    \log\left[1-p\rho\right].
    \label{eq:method_minimal_difference_entropy_density}
\end{equation}
Its stationary condition under the area constraint is
\begin{equation}
    s_p'(\rho)=\beta x.
    \label{eq:method_minimal_difference_EL}
\end{equation}
We build this relation into the neural representation by writing
\begin{equation}
    u_\phi(x;p)
    =
    \beta x+\delta u_\phi(x;p),
    \qquad
    \rho_\phi(x;p)
    =
    \left(s_p'\right)^{-1}
    \!\left(u_\phi(x;p)\right).
    \label{eq:method_minimal_difference_inverse_EL}
\end{equation}
The multiplier \(\beta\) is determined during each evaluation by the
discrete area constraint.  A single parameter-conditioned network is
used for \(p=1,2,3\).

The finite-grid normalization, inverse maps, and stabilizing terms are
given in Appendix~\ref{app:minimal_difference_neural_details}.  No
analytical or finite-grid reference profile is used in either training
objective.

\subsection{Integer projection and exact-action diagnostics}
\label{subsec:method_projection}

The hook-action solvers produce continuous, nonincreasing row lengths.
After optimization, these rows are projected onto the set of integer
partitions,
\begin{equation}
    \lambda^{\rm proj}_{n,\phi}\in\mathbb{Y}_n,
\end{equation}
by correcting the integer row lengths while preserving monotonicity and
the total number of boxes.

For the projected partition, the column lengths and hook lengths are
recomputed directly, without soft occupancies.  Evaluating the original
finite-size action then provides an exact-action check of the neural
relaxation and its integer projection.

The projection is performed only after neural training.  In the quartic
calculation, no discrete local refinement is applied to the projected
neural partition.  The neural profile and the finite-size MAP profile
therefore come from separate optimizations.  The projection procedure and
the exact-action formulas are given in
Appendix~\ref{app:projection_details}.

\subsection{Discrete MAP search and cross-\(n\) continuation}
\label{subsec:method_map_continuation}

For the quartically deformed ensemble, the finite-size MAP diagram is
defined by
\begin{equation}
    \lambda^{\rm MAP}_{n,\theta}
    \in
    \operatorname*{arg\,min}_{\lambda\vdash n}
    S^{(1)}_{n,\theta}(\lambda),
    \label{eq:method_quartic_map}
\end{equation}
where \(S^{(1)}_{n,\theta}\) is the exact integer action in
Eq.~\eqref{eq:quartic_hook_action} with \(c_h=1\).

We approximate this minimizer by searching directly over integer
partitions.  The elementary move transfers one box between rows, or
places it in a newly created row, while preserving the total number of
boxes.  Action-lowering moves are accepted directly, while
action-increasing moves are allowed during an annealing stage.

Several independent starts are used to reduce sensitivity to the initial
partition.  The search is also continued from smaller to larger \(n\):
the lowest-action partition at one size is rescaled and included among
the initial states at the next size.  The reported MAP candidate is the
partition with the lowest exact action found over all starts.

The MAP search is used for optimization rather than equilibrium sampling.
The move proposal, annealing schedule, and continuation procedure are
described in Appendix~\ref{app:map_details}.

\subsection{Corner-transfer Metropolis--Hastings sampling}
\label{subsec:method_corner_mh}

The discrete MAP search identifies a low-action partition, but does not
by itself show that the probability mass is concentrated near the same
profile.  We therefore sample the finite-size ensemble
\begin{equation}
    \mathbb{P}_{n,\theta}(\lambda)
    \propto
    \exp\left[
        -S^{(1)}_{n,\theta}(\lambda)
    \right]
\end{equation}
with a separate corner-transfer Metropolis--Hastings chain~\cite{metropolis1953equation,hastings1970monte}.

Let \(R(\lambda)\) denote the set of removable corners of \(\lambda\).
A proposal removes one corner chosen uniformly from \(R(\lambda)\),
giving an intermediate partition \(\nu\), and then adds a box at a
uniformly selected addable corner of \(\nu\).  The resulting candidate
partition is denoted by \(\lambda'\).  The acceptance probability is
\begin{equation}
    p_{\rm acc}(\lambda\rightarrow\lambda')
    =
    \min\left\{
        1,\,
        \exp\left[
            -
            \frac{
                S^{(1)}_{n,\theta}(\lambda')
                -
                S^{(1)}_{n,\theta}(\lambda)
            }{T}
        \right]
        \frac{|R(\lambda)|}{|R(\lambda')|}
    \right\}.
    \label{eq:method_corner_mh_acceptance}
\end{equation}
The ratio of removable-corner counts is the Hastings correction for the
asymmetric proposal.  The production calculations use \(T=1\), for which
the stationary distribution is the ensemble defined in
Eq.~\eqref{eq:quartic_hook_measure} with \(c_h=1\).

After burn-in and thinning, the sampled mean profile and mean action are
compared with the corresponding MAP candidate.  Agreement of the mean
profile with the MAP profile tests whether the low-action diagram lies in
the same saddle region as the typical sampled configurations.  The mean
sampled action is expected to lie slightly above the lowest action found
by the MAP search.

The corner-transfer chain is used only for equilibrium sampling and is
separate from the annealed row-transfer search used for optimization.
Chain lengths, initial states, thinning intervals, and recorded
diagnostics are given in Appendix~\ref{app:mcmc_details}.

\clearpage
\section{Benchmark recovery and scaling diagnostics}
\label{sec:benchmark_results}

We first test the variational framework on ensembles with known
analytical or asymptotic behavior.  These reference results are used
\textbf{\textit{only}} after optimization: they \textbf{\textit{do not}} enter the training objectives or
the checkpoint selection.

\subsection{Plancherel ensemble: recovery from the nonlocal hook action}
\label{subsec:results_plancherel}

The Plancherel ensemble provides a direct benchmark for the hook-based
solver.  Its finite-size action retains the nonlocal coupling between
row and column lengths, while the limiting VKLS profile is known
analytically.  We optimize the differentiable soft-hook action at
\(n=10^{5}\), without using the VKLS curve during training.  The numerical
settings are given in Appendix~\ref{app:plancherel_neural_details}.

Figure~\ref{fig:plancherel_recovery} compares the resulting neural
profile with the VKLS limit shape in the balanced English coordinates.
We measure their difference by
\begin{equation}
    E_{L^2}^{\rm Pl}
    =
    \left[
    \int_{0}^{2}
    \left|
        f_{\rm NN}(x)-f_{\rm VKLS}(x)
    \right|^2
    \,\du x
    \right]^{1/2},
    \label{eq:plancherel_absolute_l2}
\end{equation}
and obtain
\begin{equation}
    E_{L^2}^{\rm Pl}
    =
    1.0669\times10^{-2}.
    \label{eq:plancherel_l2_result}
\end{equation}
The two curves agree closely over the bulk of the support.  The largest
pointwise discrepancy occurs near \(x=0\), where the first few row
lengths are most sensitive to finite-size and soft-boundary effects.

\begin{figure}[t]
    \centering
    \includegraphics[width=0.48\textwidth]
    {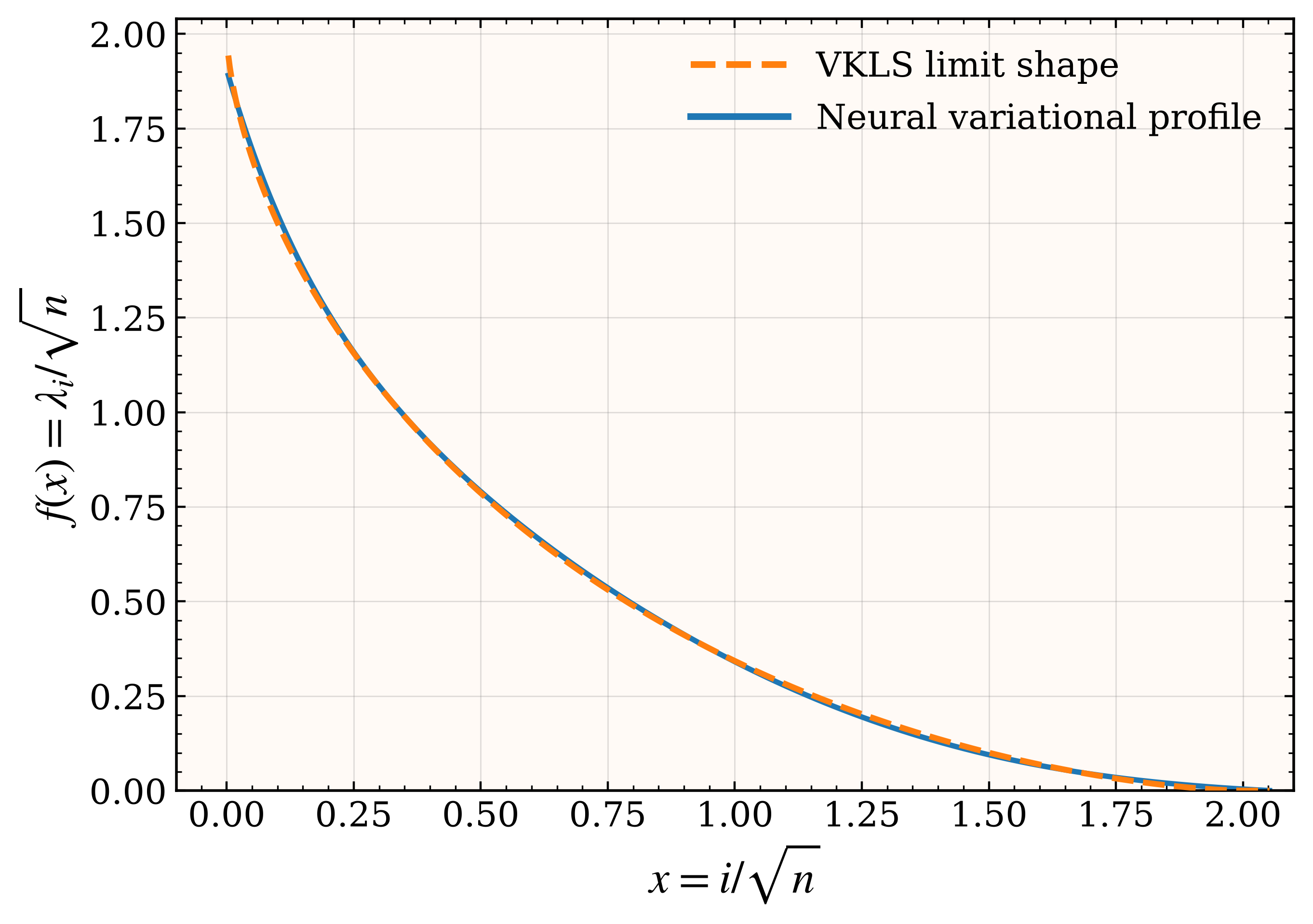}
    \hfill
    \includegraphics[width=0.48\textwidth]
    {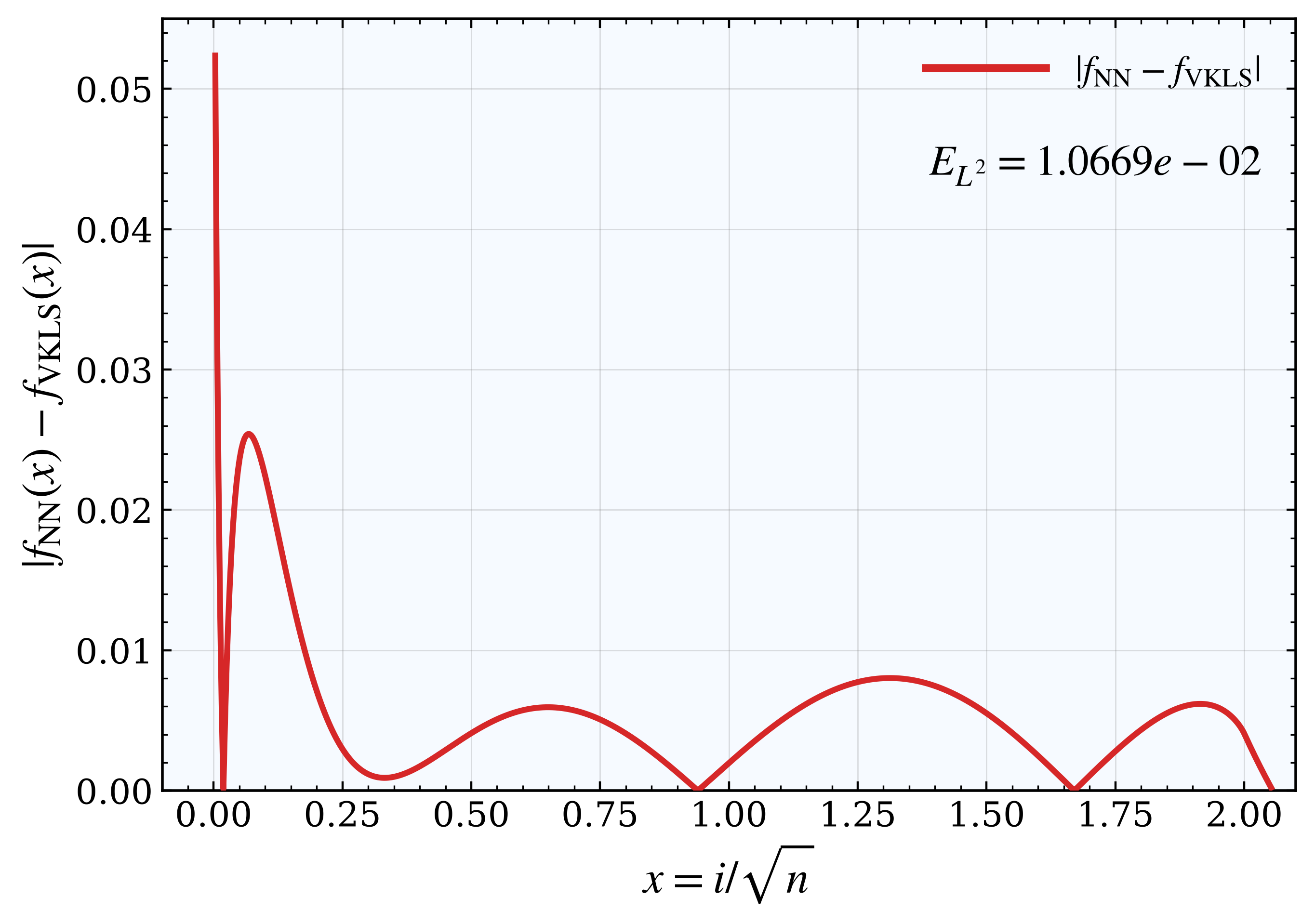}
    \caption{
    Recovery of the Plancherel limit shape from the finite-size
    soft-hook action at \(n=10^{5}\).
    Left: neural variational profile and analytical VKLS reference.
    The reference curve is used only after training.
    Right: pointwise discrepancy
    \(\lvert f_{\rm NN}(x)-f_{\rm VKLS}(x)\rvert\).
    The absolute \(L^2\) discrepancy is
    \(E_{L^2}^{\rm Pl}=1.0669\times10^{-2}\).
    }
    \label{fig:plancherel_recovery}
\end{figure}

After projection to an integer partition, the scaled largest row and
the scaled number of nonzero rows are
\begin{equation}
    \frac{\lambda_1}{\sqrt n}=1.8910,
    \qquad
    \frac{\ell(\lambda)}{\sqrt n}=2.0112.
    \label{eq:plancherel_edge_observables}
\end{equation}
Both are close to the VKLS edge value \(2\).  The projected partition
contains exactly \(10^{5}\) boxes.

The temperature continuation also improves the profile-level agreement.
The discrepancy decreases from
\(3.1622\times10^{-2}\) after the \(\tau=0.8\) stage to
the value in Eq.~\eqref{eq:plancherel_l2_result} after continuation to
\(\tau=0.4\).  Thus the nonlocal soft-hook objective recovers the known
Plancherel profile without analytical profile labels.


\clearpage
\subsection{Entropy-dominated limit shapes}
\label{subsec:results_entropy}

For uniformly weighted partition ensembles, the probability of an
individual diagram does not select a macroscopic profile.  The finite-size
MAP problem is degenerate, and the typical shape is instead determined by
the entropy of the microscopic partitions associated with a given
profile.

\subsubsection{Uniform random partitions}
\label{subsubsec:results_uniform}

We first consider ordinary uniform partitions.  The neural density is
optimized using the finite-grid Bose-type entropy introduced in
Sec.~\ref{subsec:method_entropy}.  Neither the classical limit shape nor
the finite-grid entropy maximizer is used during training or checkpoint
selection.

The calculation is performed at \(n=10^{5}\).  The grid begins at
\(x_{\min}=1/\sqrt n\), rather than at \(x=0\), because the continuum
profile has a logarithmic singularity at the left endpoint.  The
classical entropy maximizer is~\cite{vershik1996statistical}
\begin{equation}
    f_{\rm cont}(x)
    =
    -\frac{1}{c}
    \log\left(1-e^{-cx}\right),
    \qquad
    c=\frac{\pi}{\sqrt6}.
    \label{eq:uniform_continuum_profile}
\end{equation}

A direct comparison with Eq.~\eqref{eq:uniform_continuum_profile} also
contains finite-grid effects.  We therefore construct a second reference
by evaluating the Bose stationary density on the same grid, fixing its
multiplier with the discrete area constraint, and reconstructing the
boundary with the same tail sum as in the neural calculation.  We denote
this grid-consistent profile by \(f_{\rm grid}\).  Its construction and
the numerical grid are given in
Appendix~\ref{app:uniform_neural_details}.

Figure~\ref{fig:uniform_recovery} compares the neural result with both
references.  The neural profile follows the finite-grid entropy maximizer
throughout the plotted domain.  Its difference from the continuum curve
is larger near the first few grid points, where the lower cutoff and the
discrete reconstruction are most visible.

We quantify the comparison by
\begin{equation}
    E_{L^2}^{r,{\rm cut}}
    =
    \left[
    \int_{x_{\rm cut}}^{X_{\max}}
    \left|
        f_{\rm NN}(x)-f_r(x)
    \right|^2
    \,\du x
    \right]^{1/2},
    \qquad
    r\in\{{\rm grid},{\rm cont}\},
    \label{eq:uniform_cut_l2}
\end{equation}
using \(x_{\rm cut}=0.02\).  The resulting discrepancies are
\begin{equation}
    E_{L^2}^{{\rm grid},{\rm cut}}
    =
    9.8497\times10^{-4},
    \qquad
    E_{L^2}^{{\rm cont},{\rm cut}}
    =
    8.3313\times10^{-3}.
    \label{eq:uniform_l2_results}
\end{equation}
The full-domain discrepancy from the finite-grid reference is
\(1.3038\times10^{-3}\), so the agreement does not depend sensitively on
removing the endpoint region.

\begin{figure}[t]
    \centering
    \includegraphics[width=0.48\textwidth]
    {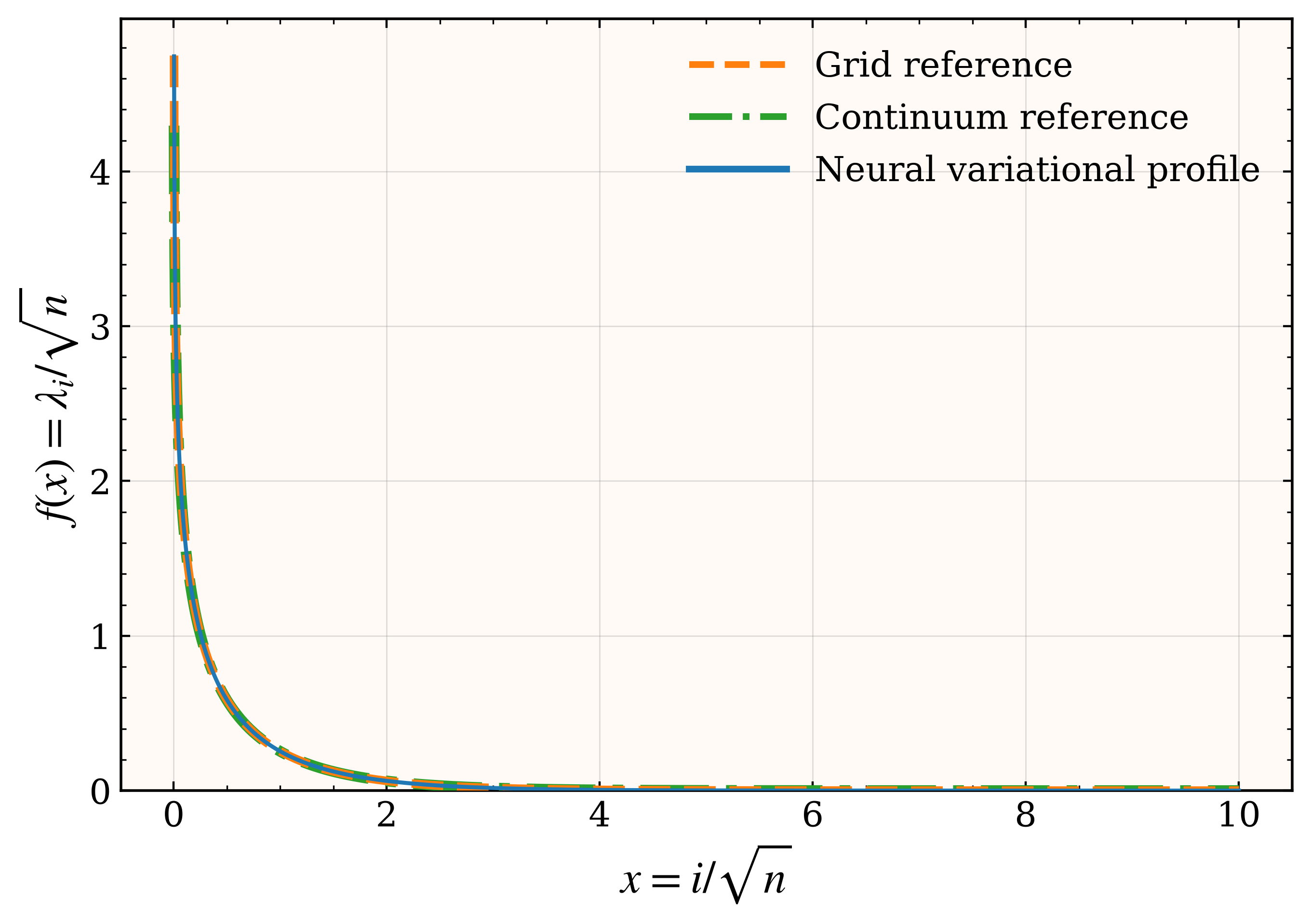}
    \hfill
    \includegraphics[width=0.48\textwidth]
    {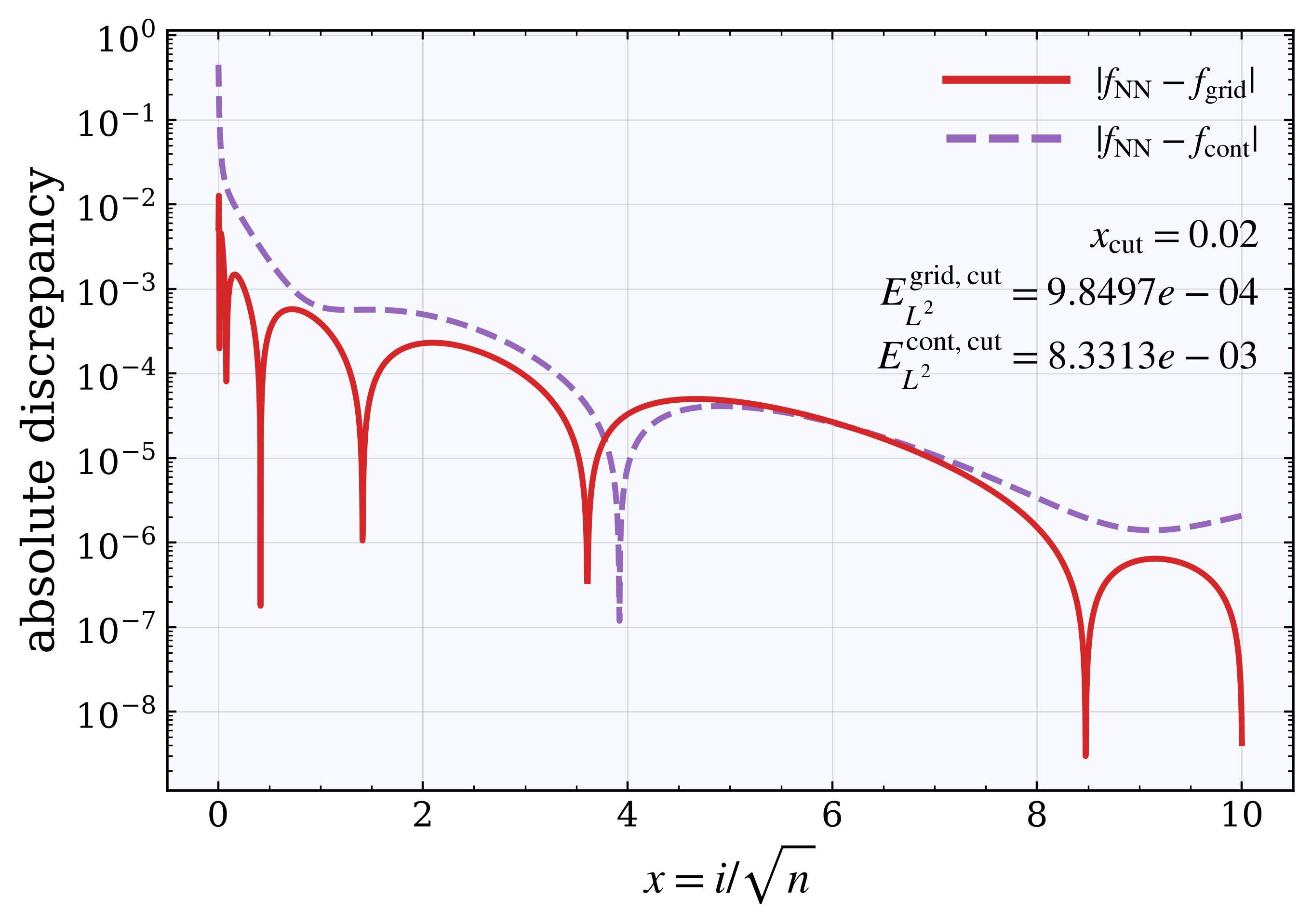}
    \caption{
    Entropy-based recovery of the uniform random-partition profile at
    \(n=10^{5}\).
    Left: neural profile, grid-consistent entropy maximizer, and
    continuum limit shape.
    Right: pointwise discrepancies from the two references.
    On \(x\geq0.02\), the corresponding absolute \(L^2\) discrepancies
    are \(9.8497\times10^{-4}\) and
    \(8.3313\times10^{-3}\).
    }
    \label{fig:uniform_recovery}
\end{figure}

The larger continuum discrepancy is concentrated near the logarithmic
endpoint and mainly reflects the difference between the continuum and
finite-grid problems.  Relative to the variational problem actually
solved on the grid, the neural discrepancy is already of order
\(10^{-3}\).  The density-based solver therefore recovers the
entropy-selected typical profile even though all individual partitions
have the same finite-size probability.


\subsubsection{Minimal-difference partitions}
\label{subsubsec:results_minimal_difference}

We next consider minimal-difference partitions with
\(p=1,2,3\), following the ordinary uniform ensemble as the \(p=0\)
case.  The constraint
\(\lambda_i-\lambda_{i+1}\geq p\) introduces an increasing degree of
exclusion between neighboring row lengths; \(p=1\) corresponds to
partitions into distinct parts.

A single parameter-conditioned network is trained simultaneously for
all three values of \(p\) using the inverse Euler--Lagrange
representation introduced in Sec.~\ref{subsec:method_entropy}.  The
finite-grid reference profiles are evaluated only after training and do
not enter either the objective or the checkpoint selection.

For each \(p\), we construct a grid-consistent stationary reference from
\begin{equation}
    \rho_{p,{\rm grid}}(x_k)
    =
    \left(s_p'\right)^{-1}
    \left(\beta_{p,{\rm grid}}x_k\right),
    \qquad
    f_{p,{\rm grid}}(x_k)
    =
    \Delta x
    \sum_{\ell=k}^{K}
    \rho_{p,{\rm grid}}(x_\ell),
    \label{eq:minimal_difference_grid_reference}
\end{equation}
where \(\beta_{p,{\rm grid}}\) is fixed by the discrete area constraint.
The reference therefore uses the same grid and tail-sum reconstruction
as the neural calculation.  Further numerical details are given in
Appendix~\ref{app:minimal_difference_neural_details}.

\begin{figure}[t]
    \centering
    \includegraphics[width=0.48\textwidth]
    {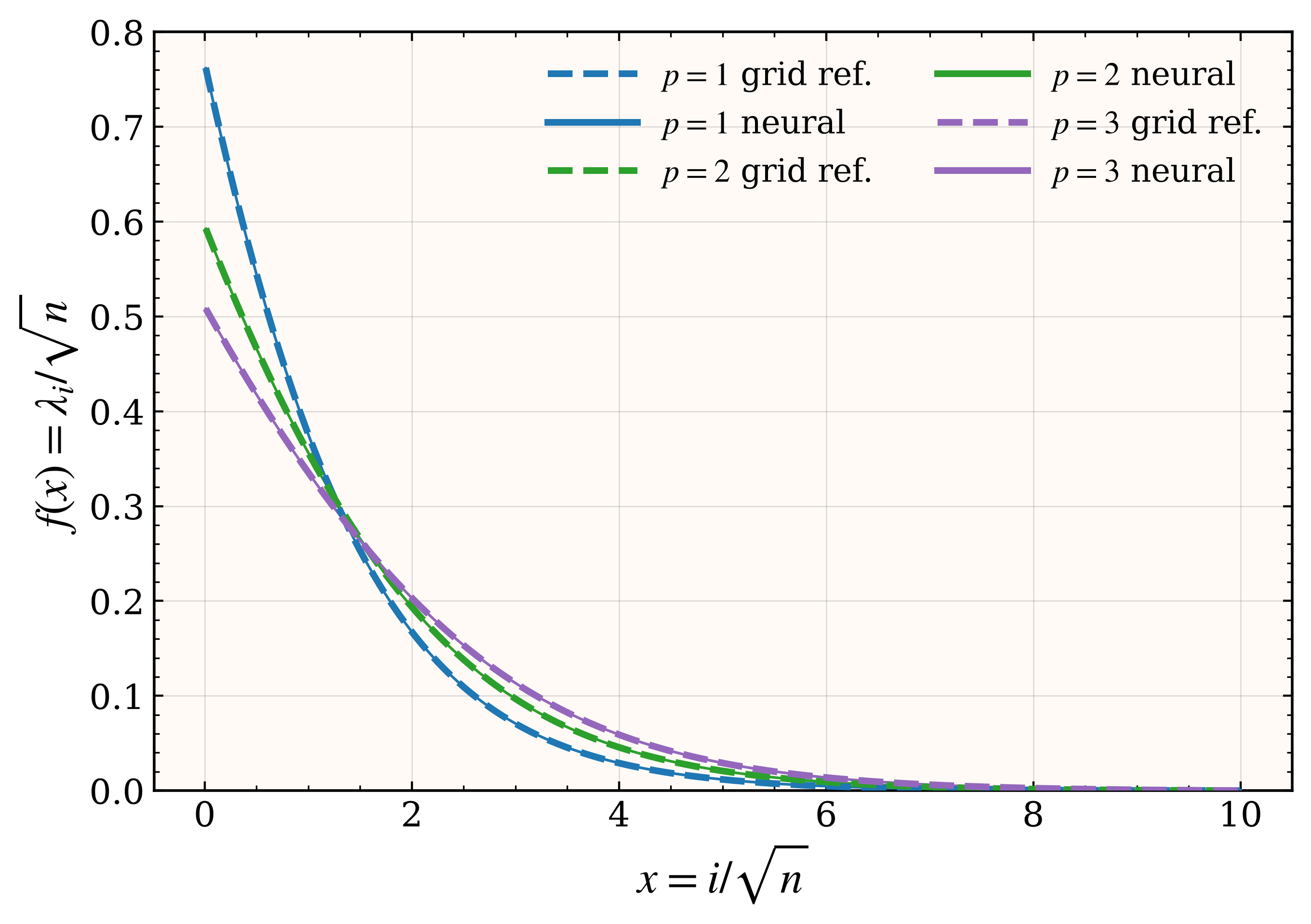}
    \hfill
    \includegraphics[width=0.48\textwidth]
    {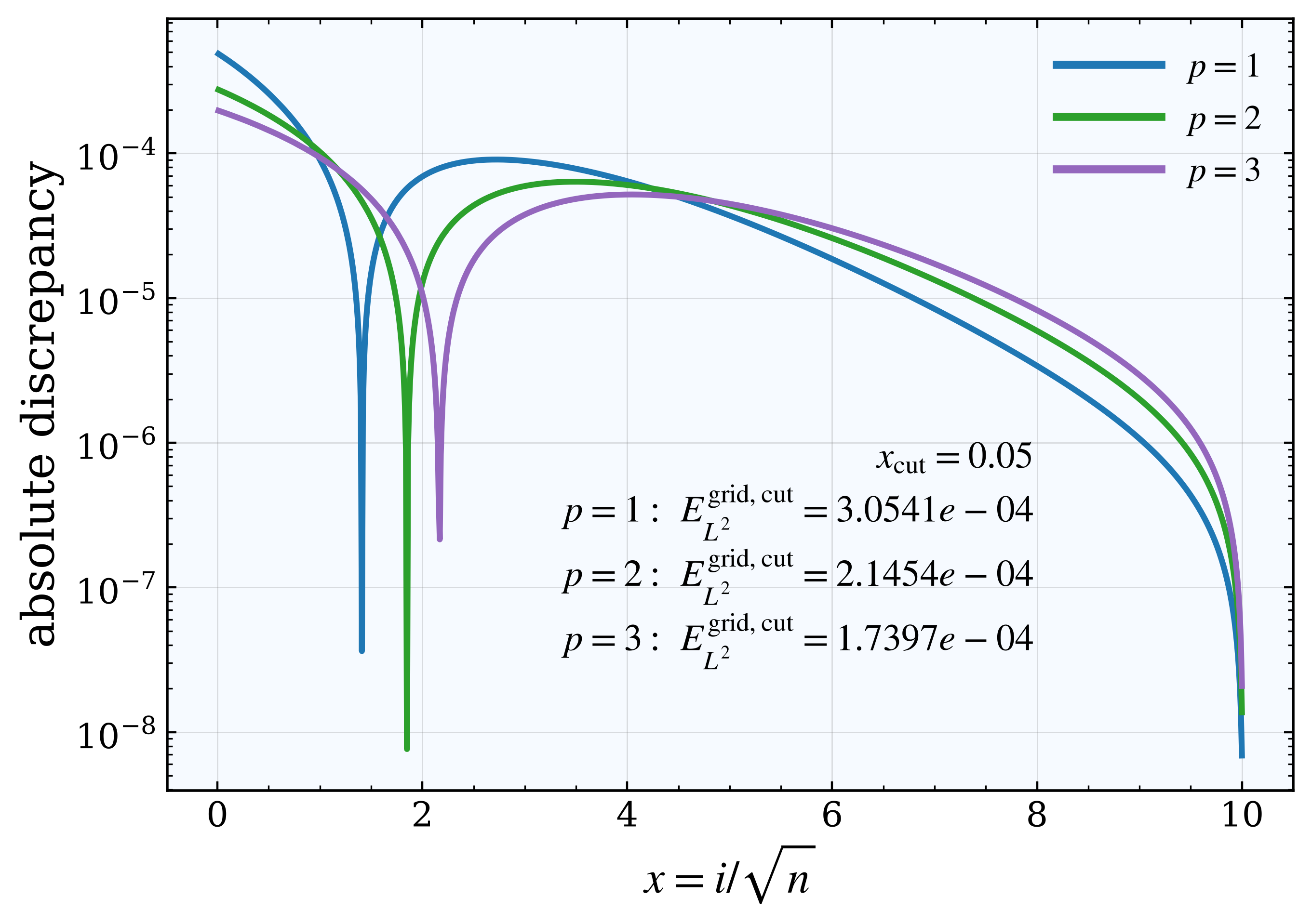}
    \caption{
    Minimal-difference partition profiles at \(n=10^{4}\) for
    \(p=1,2,3\).
    Left: neural profiles and the corresponding finite-grid
    Euler--Lagrange references.  A single parameter-conditioned network
    is used for all three values of \(p\).
    Right: pointwise absolute discrepancies on a logarithmic scale.
    On \(x\geq0.05\), the absolute \(L^2\) discrepancies are
    \(3.0541\times10^{-4}\), \(2.1454\times10^{-4}\), and
    \(1.7397\times10^{-4}\), respectively.
    }
    \label{fig:minimal_difference_recovery}
\end{figure}

The calculations are performed at \(n=10^{4}\).  As shown in
Fig.~\ref{fig:minimal_difference_recovery}, the neural and finite-grid
profiles are nearly indistinguishable for all three values of \(p\).
Increasing \(p\) lowers the profile near the left endpoint and extends
it toward larger \(x\).  This follows from the exclusion-statistics
bound
\begin{equation}
    0\leq\rho_p(x)<\frac{1}{p}.
    \label{eq:minimal_difference_density_bound}
\end{equation}
A stronger exclusion constraint reduces the allowed local density, so
the fixed area is distributed over a broader range of row indices.

We quantify the agreement using
\begin{equation}
    E_{L^2}^{p,{\rm grid},{\rm cut}}
    =
    \left[
    \Delta x
    \sum_{x_k\geq x_{\rm cut}}
    \left|
        f_{p,{\rm NN}}(x_k)
        -
        f_{p,{\rm grid}}(x_k)
    \right|^2
    \right]^{1/2},
    \qquad
    x_{\rm cut}=0.05.
    \label{eq:minimal_difference_cut_l2}
\end{equation}
For \(p=1,2,\) and \(3\), respectively, we obtain
\begin{equation}
    E_{L^2}^{p,{\rm grid},{\rm cut}}
    =
    3.0541\times10^{-4},
    \quad
    2.1454\times10^{-4},
    \quad
    1.7397\times10^{-4}.
    \label{eq:minimal_difference_l2_results}
\end{equation}
The corresponding full-domain discrepancies are all below
\(3.3\times10^{-4}\), so the agreement is not a consequence of removing
the first few grid cells.

Thus one conditioned neural representation recovers the three
exclusion-statistics profiles with errors of order \(10^{-4}\), while
also reproducing their systematic broadening with increasing \(p\).


\subsection{\(q\)-Plancherel ensemble: a scaling stress test}
\label{subsec:results_qplancherel}

The fixed-\(q\) \(q\)-Plancherel ensemble provides a test of the scaling
assumed in the neural representation.  For \(0<q<1\) fixed as
\(n\to\infty\), the leading rows are of order \(n\), rather than
\(O(\sqrt n)\).  The appropriate variables are therefore the discrete
row fractions
\begin{equation}
    \mu_i^{(n,q)}
    =
    \frac{\lambda_i}{n},
    \qquad
    i=1,2,\ldots .
    \label{eq:qplancherel_row_fractions}
\end{equation}
For each fixed row index,
\begin{equation}
    \mu_i^{(n,q)}
    \longrightarrow
    (1-q)q^{i-1},
    \qquad
    n\to\infty .
    \label{eq:qplancherel_geometric_limit}
\end{equation}
The geometric sequence is used only for post-training validation. The geometric law describes the typical asymptotics of random
\(q\)-Plancherel diagrams, whereas the neural calculation minimizes a
relaxed finite-size negative log weight.  We therefore use it here as
an external consistency benchmark; the comparison does not by itself
establish a general equivalence between the MAP and typical
\(q\)-Plancherel profiles.

We perform the calculations at
\begin{equation}
    n=5\times10^{4},
    \qquad
    q=0.2,\ 0.4,\ 0.6,\ 0.8,\ 0.95 .
\end{equation}
A fixed macroscopic row would collapse toward \(x=0\) in the balanced
coordinate \(x=i/\sqrt n\).  We therefore plot \(\lambda_i/n\) directly
against the discrete row index \(i\).

\begin{figure*}[p]
    \centering

    \includegraphics[width=0.48\textwidth]
    {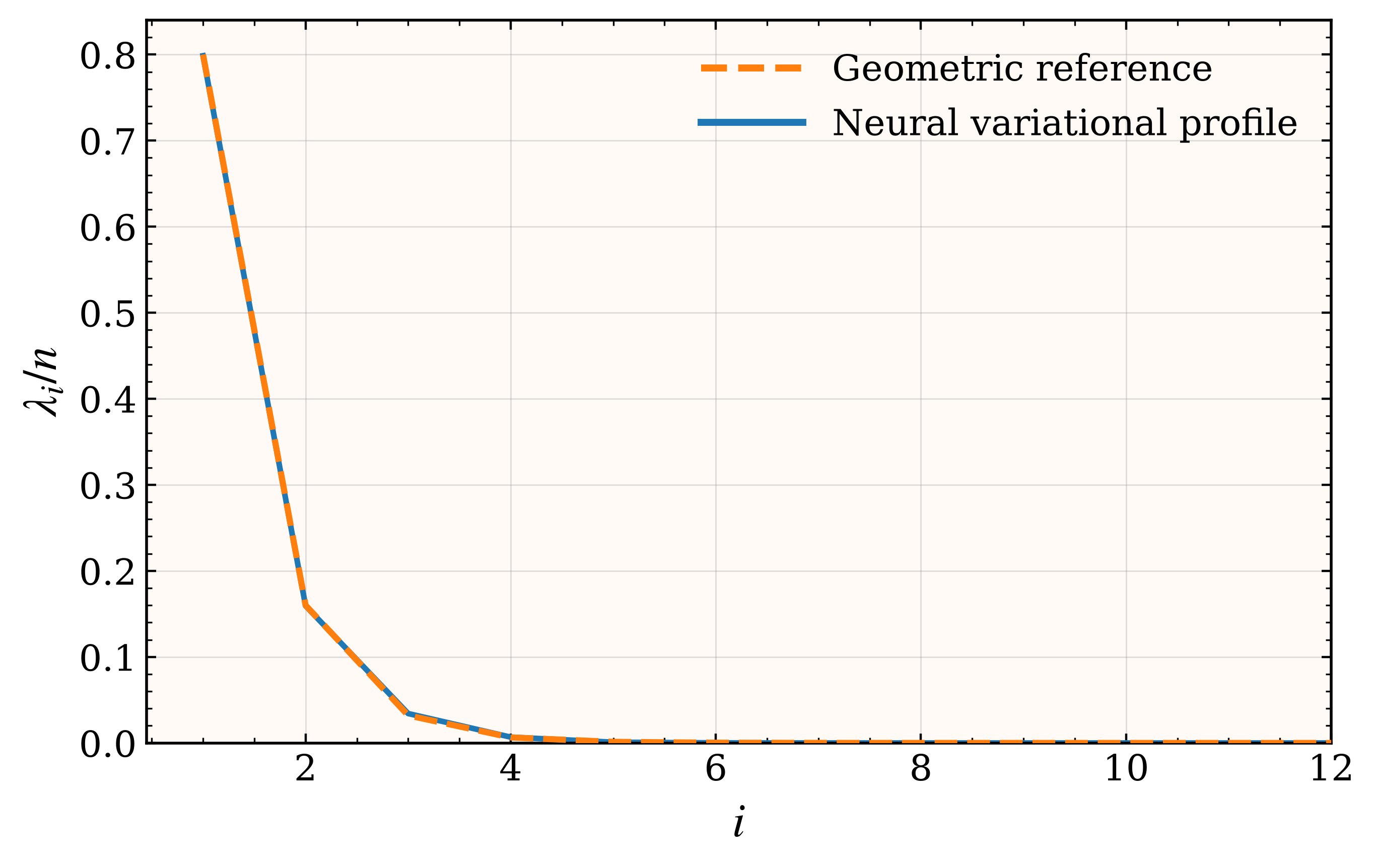}
    \hfill
    \includegraphics[width=0.48\textwidth]
    {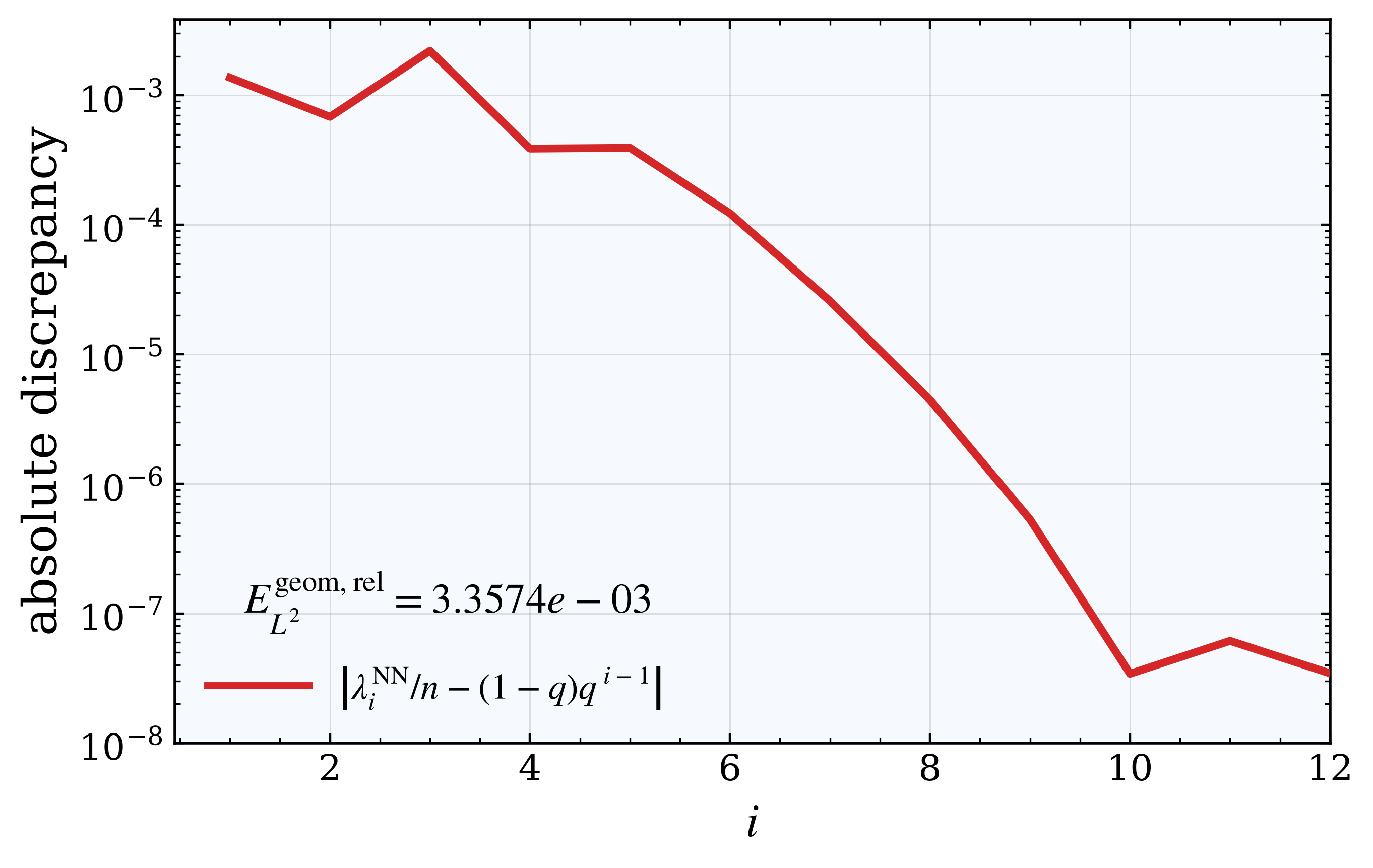}

    \vspace{0.35em}

    \includegraphics[width=0.48\textwidth]
    {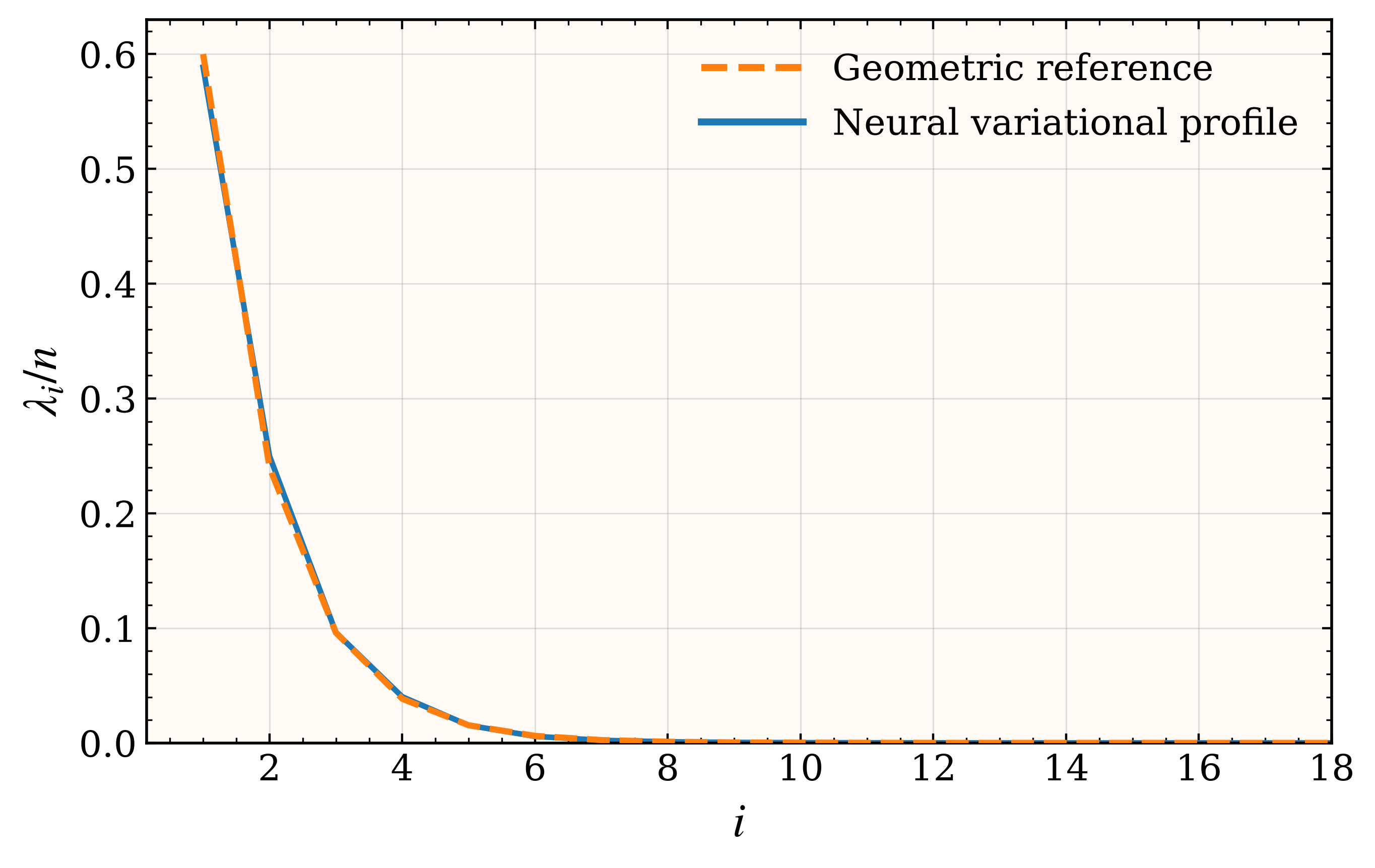}
    \hfill
    \includegraphics[width=0.48\textwidth]
    {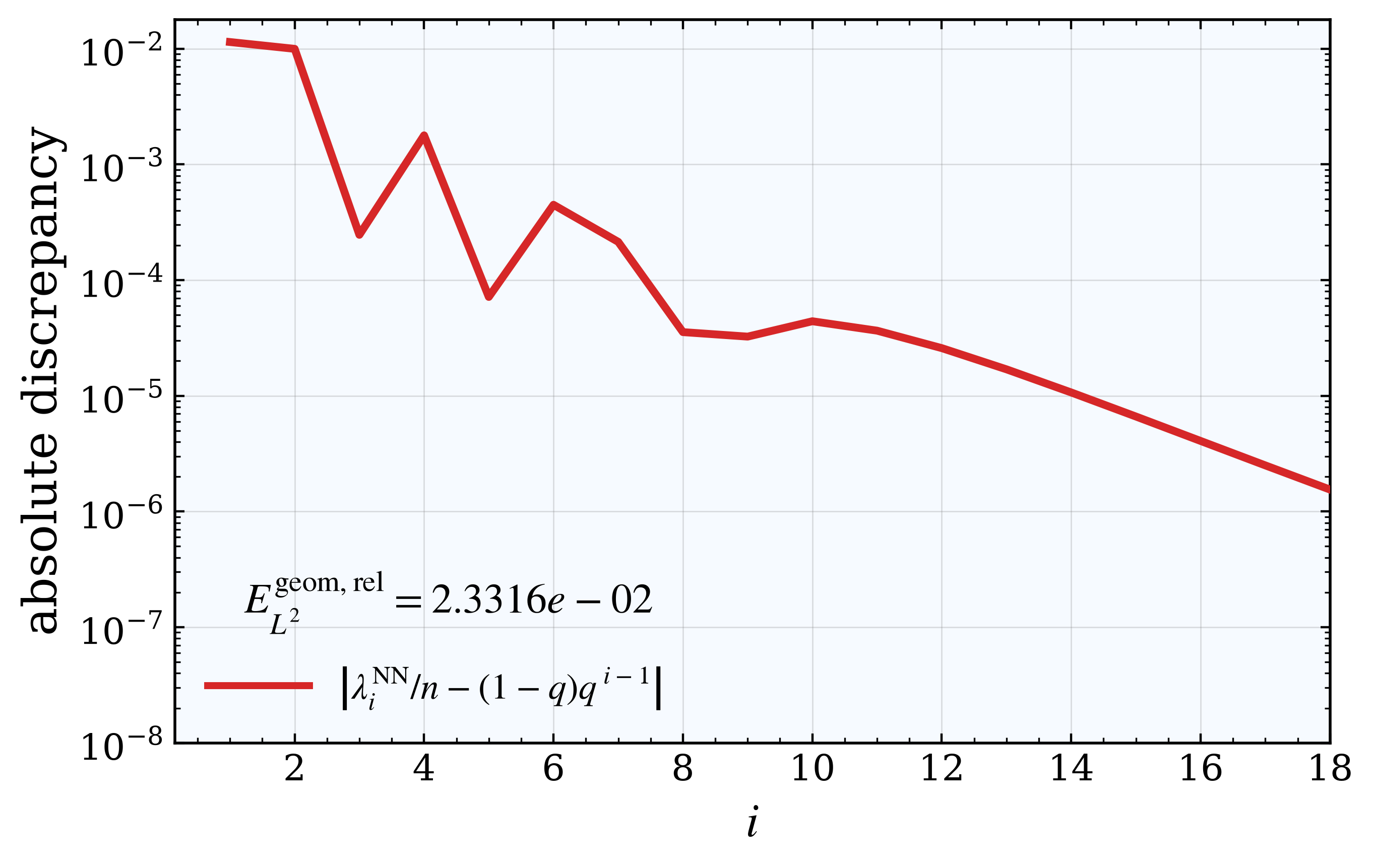}

    \vspace{0.35em}

    \includegraphics[width=0.48\textwidth]
    {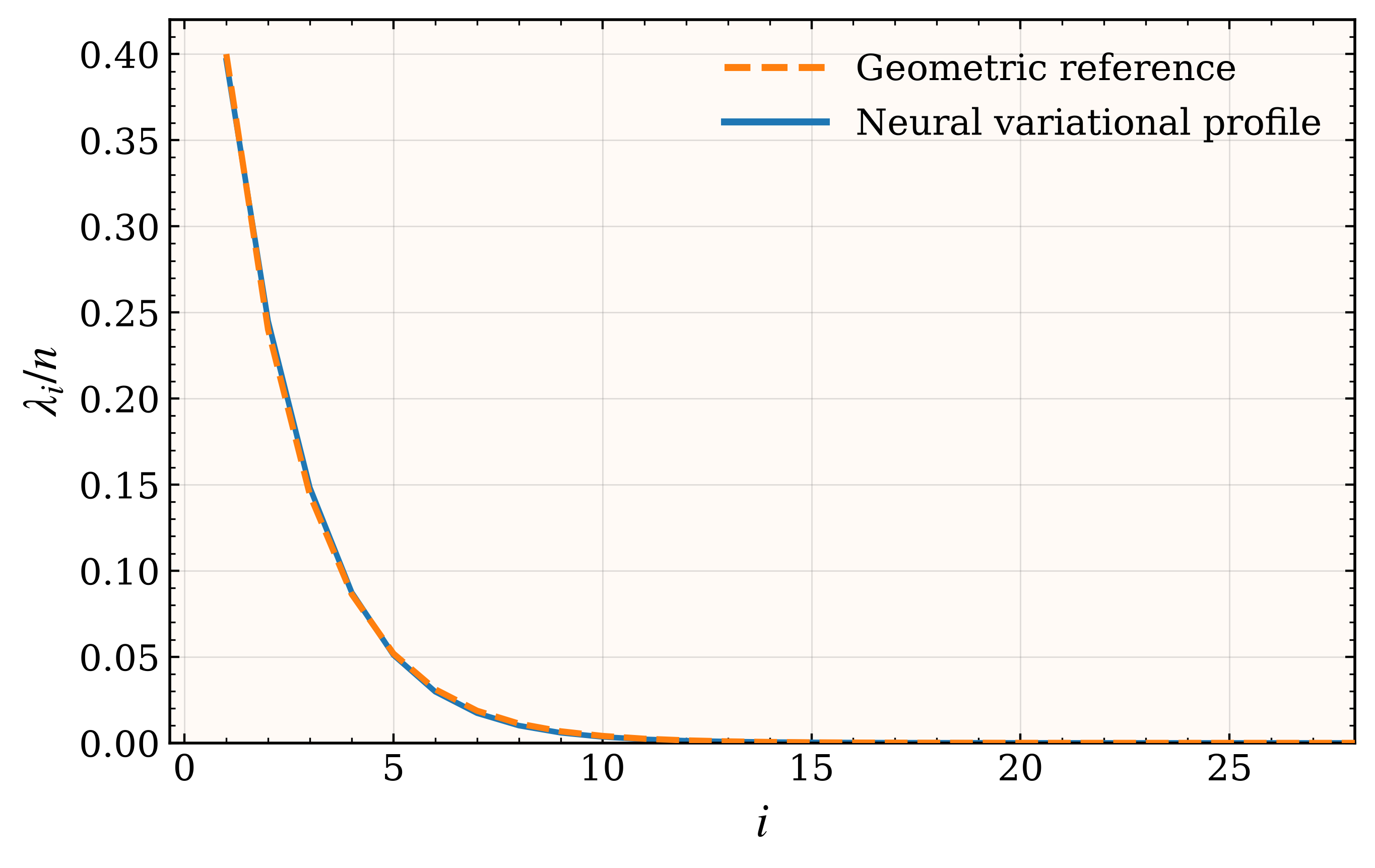}
    \hfill
    \includegraphics[width=0.48\textwidth]
    {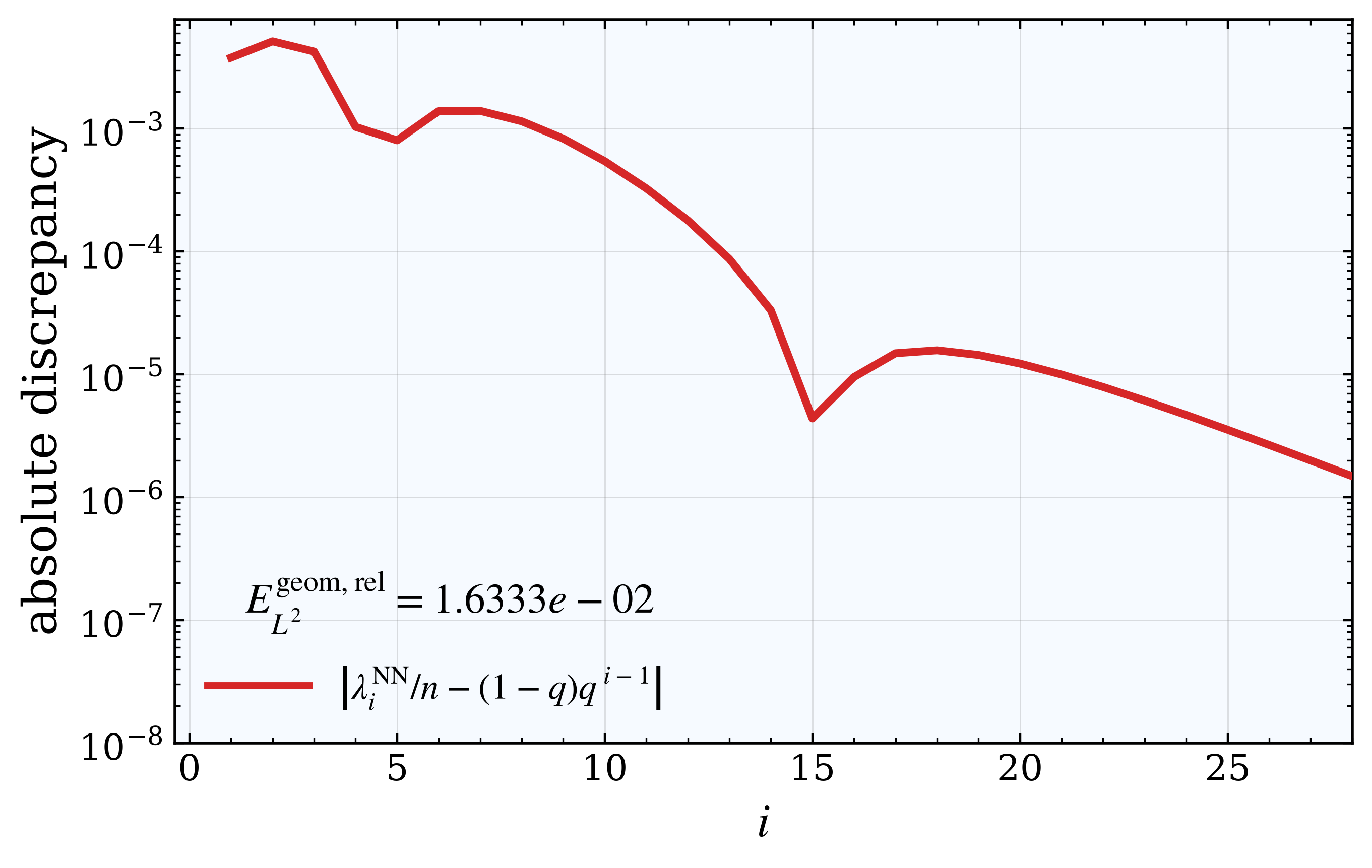}

    \vspace{0.35em}

    \includegraphics[width=0.48\textwidth]
    {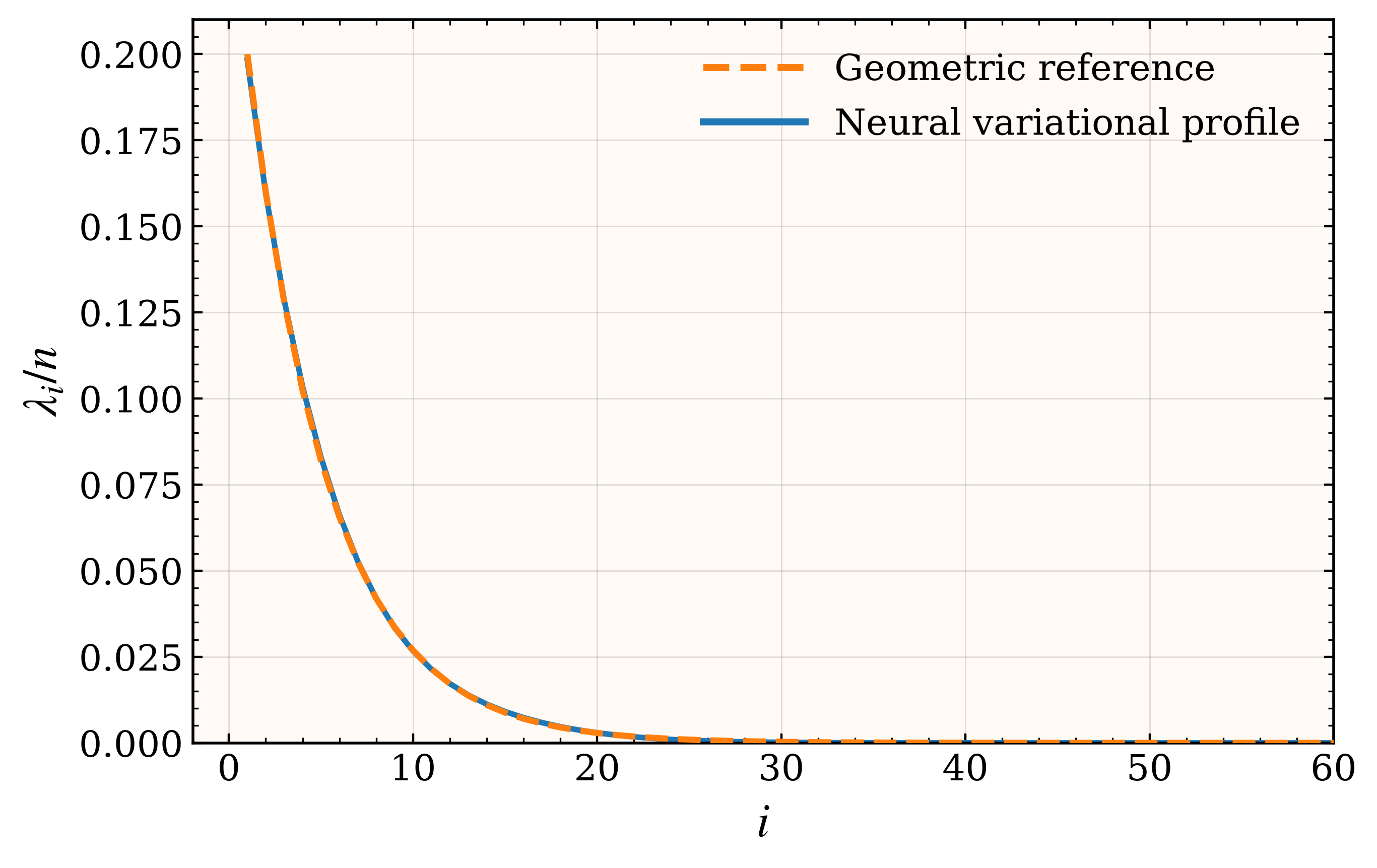}
    \hfill
    \includegraphics[width=0.48\textwidth]
    {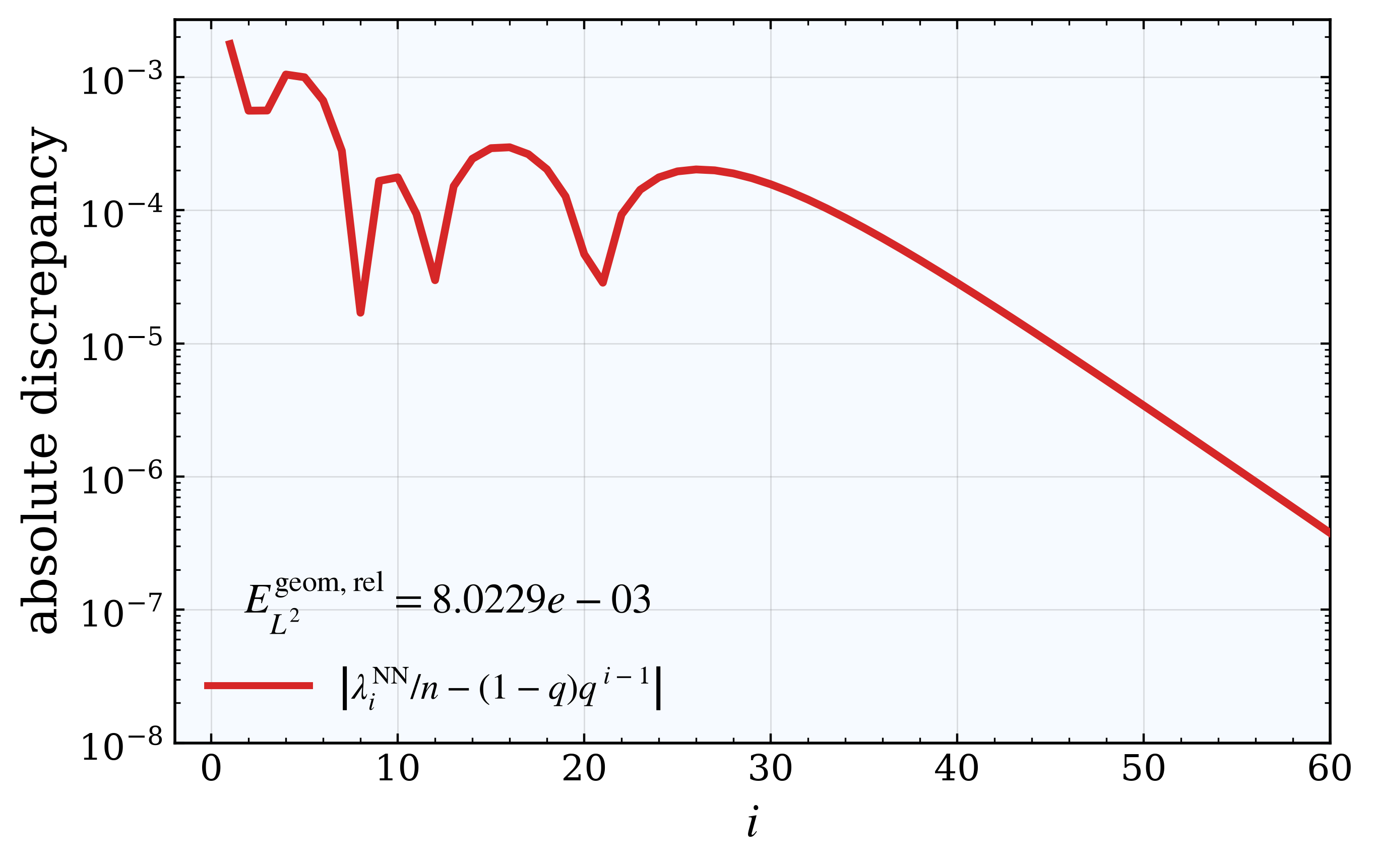}

    \caption{
    Fixed-\(q\) \(q\)-Plancherel profiles at
    \(n=5\times10^{4}\) for \(q=0.2\), \(0.4\), \(0.6\), and \(0.8\),
    from top to bottom.
    Left panels compare the neural row fractions with the geometric
    asymptotics \((1-q)q^{i-1}\); right panels show the corresponding
    pointwise absolute discrepancies.
    }
    \label{fig:qplancherel_moderate_q}
\end{figure*}

Figure~\ref{fig:qplancherel_moderate_q} shows the results for
\(q=0.2,0.4,0.6,\) and \(0.8\).  The neural calculation reproduces both
the leading row fractions and the geometric decay of the subsequent
rows.  As \(q\) increases, the first-row fraction \(1-q\) decreases and
the mass is distributed over a progressively larger number of rows.

We quantify the agreement using the relative discrete \(L^2\) discrepancy
\begin{equation}
    E_{L^2}^{\mathrm{geom,rel}}(q)
    =
    \left[
    \frac{
        \displaystyle
        \sum_{i\in\mathcal I_q}
        \left|
            \frac{\lambda_{i,\mathrm{NN}}}{n}
            -
            (1-q)q^{i-1}
        \right|^2
    }{
        \displaystyle
        \sum_{i\in\mathcal I_q}
        \left|
            (1-q)q^{i-1}
        \right|^2
    }
    \right]^{1/2},
    \label{eq:qplancherel_relative_l2}
\end{equation}
where the retained row set is
\begin{equation}
    \mathcal I_q
    =
    \left\{
        i\in\{1,\ldots,M\}:
        (1-q)q^{i-1}>10^{-12}
    \right\},
    \label{eq:qplancherel_retained_rows}
\end{equation}
and \(M\) is the number of rows in the finite neural
representation.  The cutoff removes only the numerically negligible
tail of the geometric reference.  
For \(q=0.2,0.4,0.6,0.8,\) and \(0.95\), the relative
\(L^2\) discrepancies are
\begin{equation}
\begin{array}{c|c}
    q
    &
    E_{L^2}^{\mathrm{geom,rel}}
    \\
    \hline
    0.2  & 3.3574\times10^{-3}
    \\
    0.4  & 2.3316\times10^{-2}
    \\
    0.6  & 1.6333\times10^{-2}
    \\
    0.8  & 8.0229\times10^{-3}
    \\
    0.95 & 1.9428\times10^{-2}
\end{array}
\label{eq:qplancherel_error_results}
\end{equation}
All five discrepancies remain below \(2.4\%\).

\begin{figure*}[t]
    \centering

    \includegraphics[width=0.48\textwidth]
    {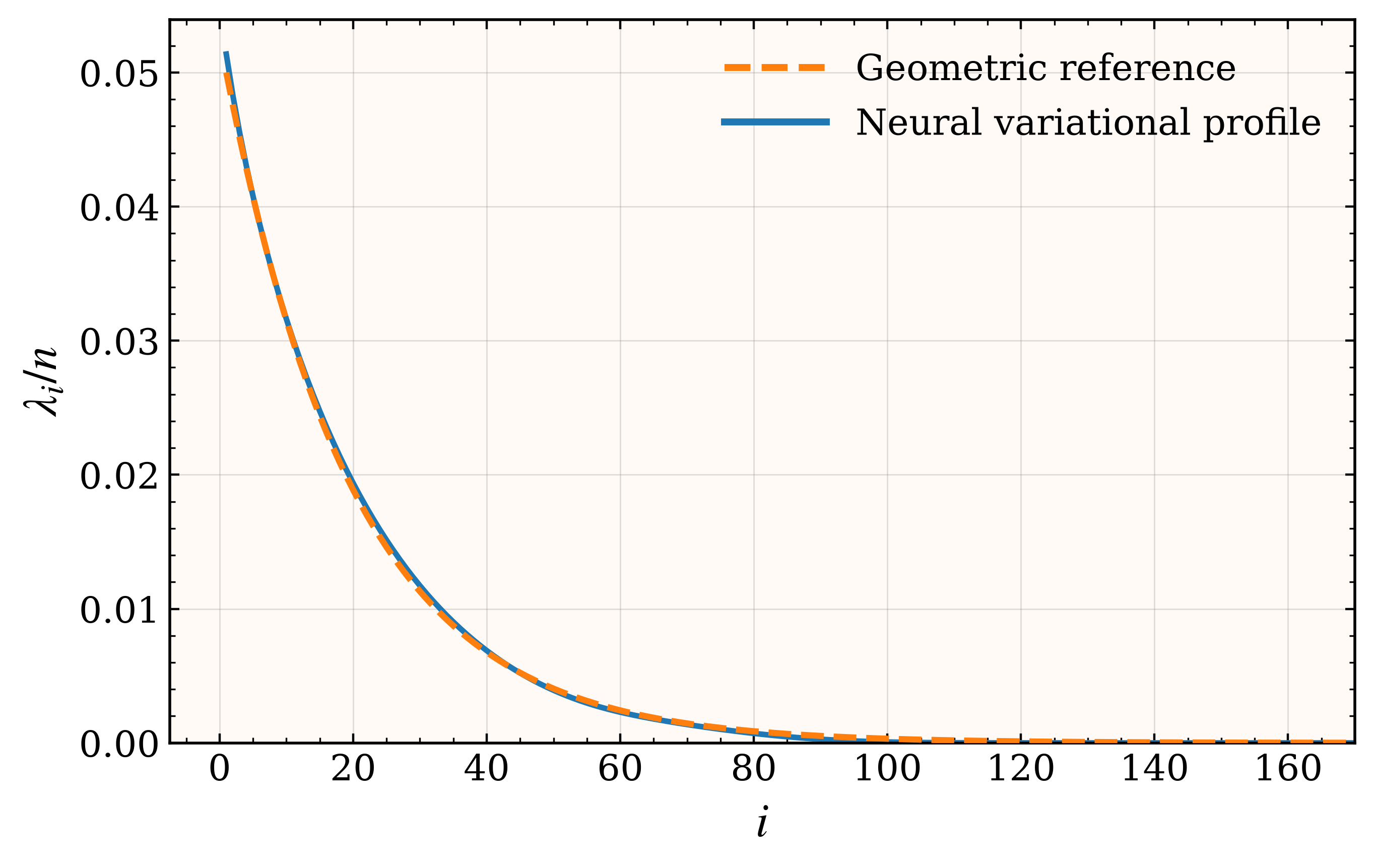}
    \hfill
    \includegraphics[width=0.48\textwidth]
    {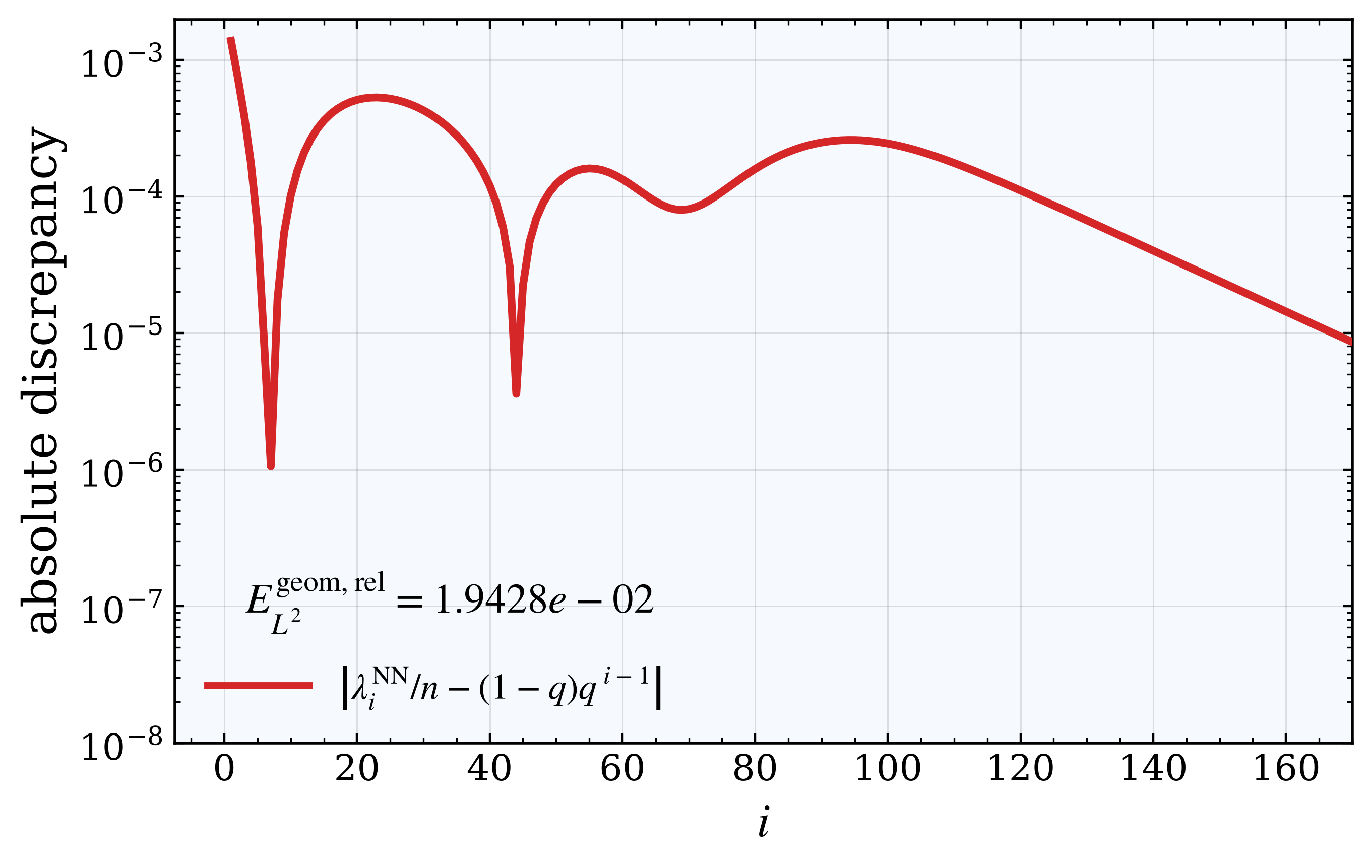}

    \caption{
    Near-unity fixed-\(q\) test at \(q=0.95\) and
    \(n=5\times10^{4}\).
    The left panel compares the neural row fractions with the geometric
    asymptotics, while the right panel shows the pointwise absolute
    discrepancy.
    }
    \label{fig:qplancherel_q095}
\end{figure*}

The near-unity case \(q=0.95\) is shown separately in
Fig.~\ref{fig:qplancherel_q095}.  Here the leading fraction is only
\(1-q=0.05\), and the slower geometric decay spreads the macroscopic
mass over many more rows.  The neural profile continues to follow the
geometric sequence through this extended tail.

The calculation therefore recovers the fixed-\(q\) \(O(n)\) row scaling
and its geometric mass distribution once the neural representation is
expressed in the appropriate variables.  This behavior is distinct from
the continuous \(O(\sqrt n)\) profiles of the preceding ensembles.

\section{Quartically deformed hook-length ensemble:
numerical saddle shapes}
\label{sec:quartic_results}

For the benchmark ensembles, the numerical profiles could be compared
with known analytical or asymptotic results.  No such reference is
assumed for the quartically deformed hook-length ensemble.  We instead
compare three separate calculations: a finite-\(n\) MAP search over
integer partitions, a neural relaxation at much larger \(n\), and
corner-transfer Metropolis--Hastings sampling of the finite-temperature
ensemble.

All calculations in this section use \(c_h=1\), with
\begin{equation}
    S^{(1)}_{n,\theta}(\lambda)
    =
    \sum_{u\in\lambda}\log h(u)
    +
    \theta\sqrt n
    \sum_{i\geq1}
    \left(
        \frac{\lambda_i}{\sqrt n}
    \right)^4 .
    \label{eq:quartic_action_results}
\end{equation}
The model is therefore a quartically deformed hook-length ensemble,
rather than the strict \(c_h=2\) deformation of the Plancherel measure.

\subsection{Undeformed control and cross-\(n\) stability}
\label{subsec:quartic_control}

At \(\theta=0\), the quartic term vanishes.  The remaining hook action
differs from the Plancherel action only by an overall positive factor,
which does not change its MAP diagrams.  The VKLS profile can therefore
be used as an external control for the undeformed MAP sequence.

For each value of \(\theta\), the discrete search is continued through
\begin{equation}
    n=500
    \longrightarrow
    1000
    \longrightarrow
    2000
    \longrightarrow
    4000
    \longrightarrow
    8000.
    \label{eq:quartic_continuation_sequence}
\end{equation}
The lowest-action partition found at one size is rescaled and included
among the initial states at the next size.  It is combined with
independent initial partitions in a multi-start search using the exact
integer action.  Details of the continuation and local moves are given
in Appendix~\ref{app:map_details}.

For the undeformed sequence, the scaled largest row increases from
\(1.7889\) at \(n=500\) to \(1.9118\) at \(n=8000\), while the scaled
number of nonzero rows changes from \(1.7889\) to \(1.9118\).  Over the
same sequence, the relative \(L^2\) discrepancy from the VKLS profile
decreases from \(0.2962\) to \(0.1470\).  The approach is slow at these
finite sizes, but all three diagnostics move toward the expected
undeformed values.

Table~\ref{tab:quartic_cross_n_stability} compares the principal MAP
observables at the two largest sizes for all four deformation strengths.
For \(\theta>0\), the changes from \(n=4000\) to \(n=8000\) remain below
approximately \(2\%\).

\begin{table}[t]
\caption{
Principal MAP observables at the two largest sizes in the continuation
sequence.
}
\label{tab:quartic_cross_n_stability}
\centering
\begingroup
\renewcommand{\arraystretch}{1.12}
\setlength{\tabcolsep}{8pt}
\begin{ruledtabular}
\begin{tabular}{c cc cc}
&
\multicolumn{2}{c}{\(\lambda_1/\sqrt n\)}
&
\multicolumn{2}{c}{\(\ell(\lambda)/\sqrt n\)}
\\
\(\theta\)
&
\(n=4000\)
&
\(n=8000\)
&
\(n=4000\)
&
\(n=8000\)
\\
\hline
\(0.0\) & \(1.8657\) & \(1.9118\) & \(1.8974\) & \(1.9118\) \\
\(0.1\) & \(1.3598\) & \(1.3752\) & \(2.1662\) & \(2.1913\) \\
\(0.5\) & \(1.0119\) & \(1.0174\) & \(2.6089\) & \(2.6497\) \\
\(1.0\) & \(0.8696\) & \(0.8721\) & \(2.9251\) & \(2.9516\)
\end{tabular}
\end{ruledtabular}
\endgroup
\end{table}

The undeformed control and the stability of the deformed observables
indicate that the \(n=8000\) MAP profiles are not isolated outcomes of a
single large-\(n\) search.  These finite-size results do not establish the
existence or uniqueness of a limiting profile.  They provide the
discrete baseline for the neural and MCMC comparisons below.

\subsection{Deformation-induced redistribution of boxes}
\label{subsec:quartic_deformation}

We now examine how the quartic term changes the finite-size MAP profile.
The factor \(\lambda_i^4\) penalizes long rows more strongly than short
ones.  Since the total number of boxes is fixed, suppressing the leading
rows requires the area to be redistributed over a larger number of
shorter rows.

\begin{figure*}[t]
    \centering
    \includegraphics[width=0.75\textwidth]
    {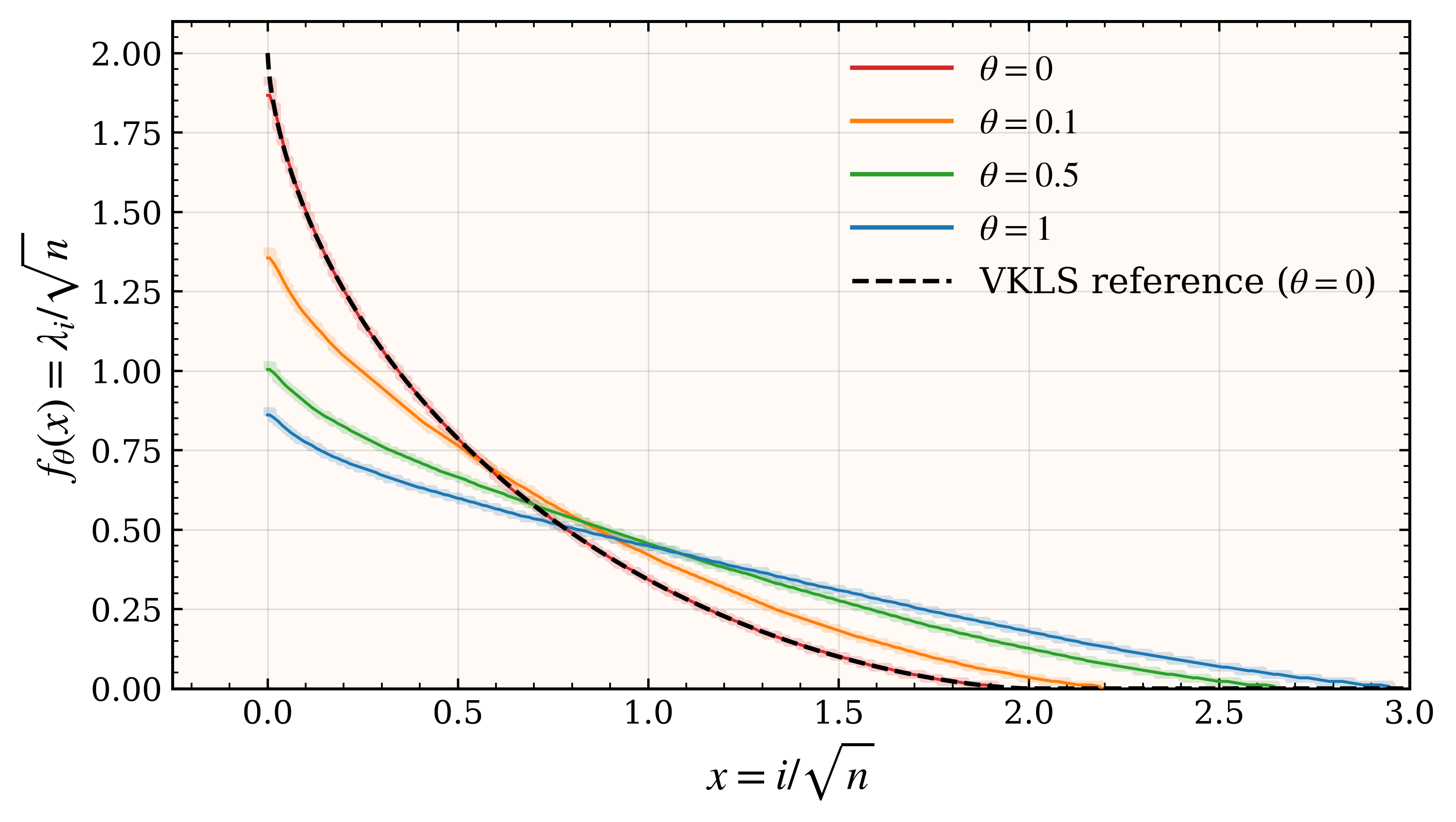}
    \caption{
    MAP profiles of the quartically deformed hook-length ensemble at
    \(n=8000\) for \(\theta=0\), \(0.1\), \(0.5\), and \(1\).
    The dashed black curve is the VKLS reference for the undeformed
    case.  Faint traces show the finite-size staircases, while the solid
    curves are coarse-grained representations used for visualization.
    }
    \label{fig:quartic_map_profiles}
\end{figure*}

Figure~\ref{fig:quartic_map_profiles} shows the MAP profiles at
\(n=8000\).  Increasing \(\theta\) lowers the profile near the left edge
and extends it toward larger \(x\).  The deformation therefore changes
the shape of the diagram rather than producing a uniform rescaling.

The change is summarized by the scaled largest row and the scaled number
of nonzero rows in Table~\ref{tab:quartic_deformation_observables}.
Across the range studied, \(\lambda_1/\sqrt n\) decreases monotonically
from \(1.9118\) to \(0.8721\), while
\(\ell(\lambda)/\sqrt n\) increases from \(1.9118\) to \(2.9516\).

\begin{table}[t]
\caption{
Principal MAP observables at \(n=8000\).
}
\label{tab:quartic_deformation_observables}
\centering
\begingroup
\renewcommand{\arraystretch}{1.12}
\setlength{\tabcolsep}{10pt}
\begin{ruledtabular}
\begin{tabular}{ccc}
\(\theta\)
&
\(\lambda_1/\sqrt n\)
&
\(\ell(\lambda)/\sqrt n\)
\\
\hline
\(0.0\) & \(1.9118\) & \(1.9118\) \\
\(0.1\) & \(1.3752\) & \(2.1913\) \\
\(0.5\) & \(1.0174\) & \(2.6497\) \\
\(1.0\) & \(0.8721\) & \(2.9516\)
\end{tabular}
\end{ruledtabular}
\endgroup
\end{table}

Already at \(\theta=0.1\), the scaled largest row is reduced by about
\(28\%\) relative to the undeformed control, while the scaled number of
rows increases by about \(15\%\).  At \(\theta=1\), the corresponding
changes are approximately \(54\%\).  Because every scaled profile has
unit area, the suppression at small \(x\) is accompanied by an
enhancement at larger \(x\), accounting for the crossings between the
curves.

The deformation thus produces the ordered trend
\begin{equation}
    \theta\uparrow
    \qquad\Longrightarrow\qquad
    \frac{\lambda_1}{\sqrt n}\downarrow,
    \qquad
    \frac{\ell(\lambda)}{\sqrt n}\uparrow.
    \label{eq:quartic_redistribution_trend}
\end{equation}

\subsection{Neural large-\(n\) profiles and agreement with the discrete MAP}
\label{subsec:quartic_neural}

We next solve the same deformed hook-length problem using the neural
relaxation introduced in Sec.~\ref{sec:method}.  The calculation is
performed at
\begin{equation}
    n=2\times10^{7},
    \qquad
    \theta=0,\ 0.1,\ 0.5,\ 1.
\end{equation}
Neither the MAP partitions nor the MCMC profiles enter the neural
objective, initialization, or checkpoint selection.  The calculation
at one value of \(\theta\) is initialized from the neural parameters
obtained at the preceding value, but no information from either
discrete calculation is used in this continuation.

\begin{figure*}[t]
    \centering
    \includegraphics[width=0.75\textwidth]
    {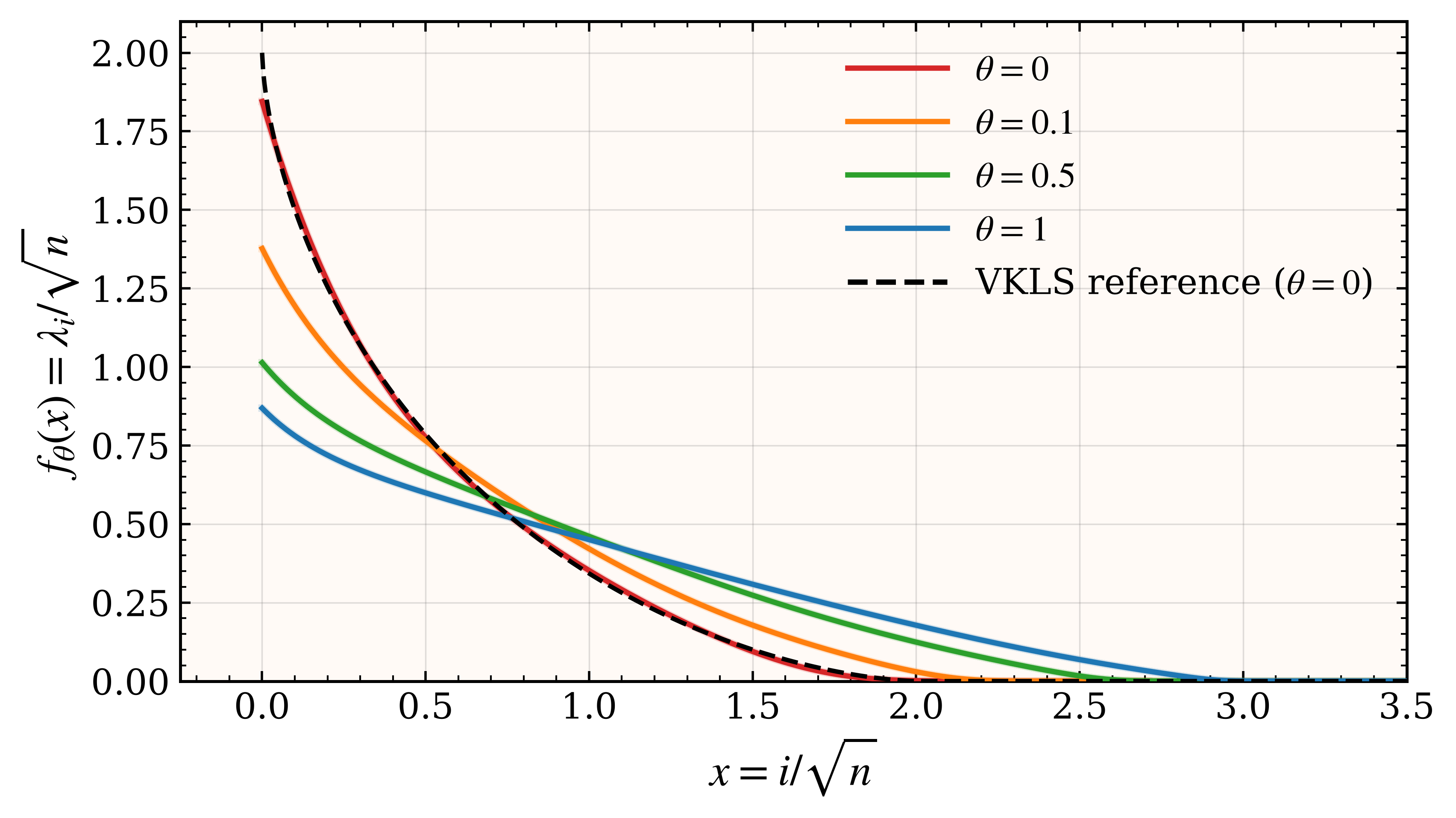}
    \caption{
    Neural profiles for \(\theta=0\), \(0.1\), \(0.5\), and \(1\) at
    \(n=2\times10^{7}\).
    Solid curves show the continuous neural outputs, while the faint
    staircases show their integer projections.
    The dashed black curve is the VKLS reference for the undeformed
    case.
    }
    \label{fig:quartic_neural_profiles}
\end{figure*}

The continuous rows are projected to integer partitions only after
training.  For all four values of \(\theta\), the relative \(L^2\)
difference between the continuous profile and its integer projection is
below \(2\times10^{-4}\).  The projection therefore has no visible
effect on the macroscopic profiles shown in
Fig.~\ref{fig:quartic_neural_profiles}.

The neural profiles reproduce the deformation trend found in the
finite-size MAP calculation.  Increasing \(\theta\) suppresses the
leading rows and shifts the redistributed area toward larger values of
\(x\).  The resulting profiles are therefore lower near the left edge
and extend over a broader range of row indices.  This behavior is
already present in the continuous relaxation and is unchanged by the
subsequent integer projection.

For \(\theta>0\), no analytical saddle profile is available.  We
therefore compare the neural result with the independently optimized
MAP profile.  The comparison is performed on the common scaled interval
\(0\leq x\leq3.5\), using
\begin{equation}
    E_{L^2}^{\rm NN/MAP}
    =
    \left[
    \frac{
        \displaystyle
        \int_{0}^{3.5}
        \left|
            f^{\rm NN}_{\theta}(x)
            -
            f^{\rm MAP}_{\theta}(x)
        \right|^2
        \,\du x
    }{
        \displaystyle
        \int_{0}^{3.5}
        \left|
            f^{\rm MAP}_{\theta}(x)
        \right|^2
        \,\du x
    }
    \right]^{1/2}.
    \label{eq:quartic_neural_map_l2}
\end{equation}
Here \(f^{\rm NN}_{\theta}\) denotes the profile obtained from the
integer projection of the neural output.  
The continuous and projected neural profiles differ by less than
\(2\times10^{-4}\) in relative \(L^2\), so this choice is immaterial at
the precision quoted here.

\begin{figure*}[t]
    \centering
    \includegraphics[width=0.80\textwidth]
    {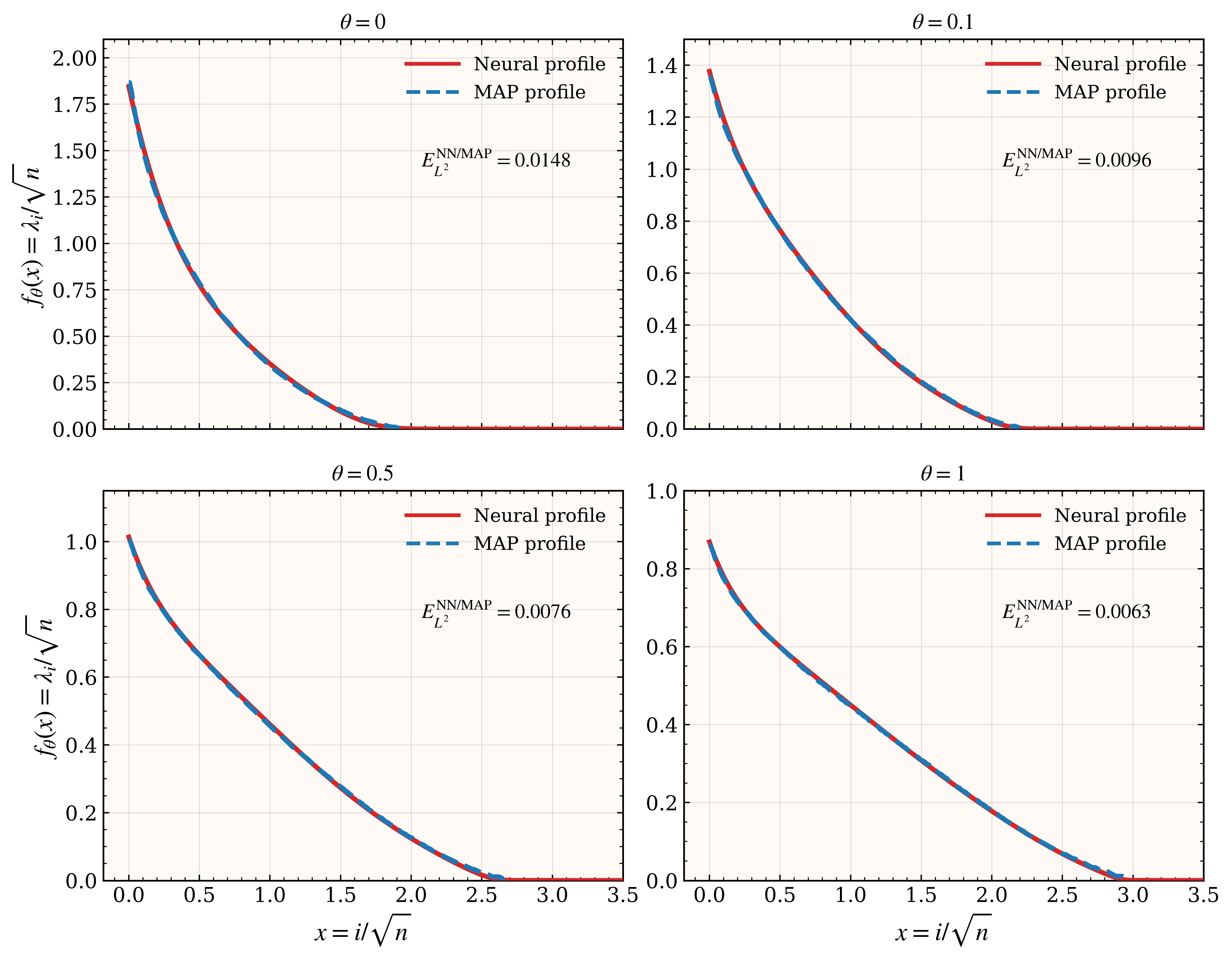}
    \caption{
    Comparison of the neural and MAP profiles for
    \(\theta=0\), \(0.1\), \(0.5\), and \(1\).
    Red solid curves show the projected neural profiles at
    \(n=2\times10^{7}\), while blue dashed curves show
    area-preserving coarse-grained representations of the MAP
    partitions at \(n=8000\).
    The annotations give the relative profile discrepancies defined in
    Eq.~\eqref{eq:quartic_neural_map_l2}.
    }
    \label{fig:quartic_neural_map}
\end{figure*}

The MAP partition is intrinsically a finite-size staircase.  For the
macroscopic comparison in Fig.~\ref{fig:quartic_neural_map}, 
we represent it using the same area-preserving coarse-graining
prescription as in Fig.~\ref{fig:quartic_map_profiles}. 
The coarse graining uses a
moving-average width \(\Delta x=0.05\), followed by a normalization that
preserves the unit area of the scaled Young diagram.  It is used only
for the profile visualization and for the discrepancy in
Eq.~\eqref{eq:quartic_neural_map_l2}.  The underlying integer partition,
its exact action, and the observables reported in
Table~\ref{tab:quartic_deformation_observables} are not modified.

Figure~\ref{fig:quartic_neural_map} compares the
\(n=2\times10^{7}\) projected neural profiles with the coarse-grained
representations of the \(n=8000\) MAP partitions.  The two calculations
are therefore compared as macroscopic profiles in their respective
\(\sqrt n\)-scaled coordinates.  This is not a comparison of
microscopic partitions at the same value of \(n\).

For \(\theta=0,0.1,0.5,\) and \(1\), respectively, the relative
profile discrepancies are
\begin{equation}
    E_{L^2}^{\rm NN/MAP}
    =
    0.0148,\qquad
    0.0096,\qquad
    0.0076,\qquad
    0.0063.
    \label{eq:quartic_neural_map_results}
\end{equation}
The discrepancy is below one percent for all three positively deformed
cases.  The undeformed control gives the largest difference,
approximately \(1.5\%\), consistent with the slower finite-size
approach of the \(\theta=0\) MAP sequence.

The leading-row observables give a comparison that does not involve
coarse graining.  For \(\theta=0.1,0.5,\) and \(1\), the neural values
of \(\lambda_1/\sqrt n\) are
\begin{equation}
    1.3768,\qquad
    1.0136,\qquad
    0.8689,
\end{equation}
respectively, while the corresponding \(n=8000\) MAP values are
\begin{equation}
    1.3752,\qquad
    1.0174,\qquad
    0.8721.
\end{equation}
The relative differences are below \(0.4\%\) in all three deformed
cases.

The agreement is not imposed by a shared optimization target.  The
neural solver minimizes a differentiable relaxation over continuous
monotone profiles, whereas the MAP calculation searches directly over
integer partitions using the exact action and cross-\(n\)
continuation.  The MAP partitions are not used during neural training,
and no exact-action local refinement is applied to the projected neural
partitions.  Their agreement therefore provides an independent
profile-level check of the deformation-dependent saddle family.

\subsection{Metropolis--Hastings validation of the MAP profiles}
\label{subsec:quartic_mcmc}

The neural relaxation and the MAP search identify closely agreeing
profiles, but neither calculation directly probes the probability mass
of the finite-size ensemble.  We therefore sample the same
\(c_h=1\) ensemble with the corner-transfer Metropolis--Hastings
algorithm described in Sec.~\ref{subsec:method_corner_mh}.  The
production chains use \(T=1\), so no additional cooling is introduced.

After burn-in and thinning, the retained partitions define the mean
profile
\(\overline f^{\,\rm MCMC}_{n,\theta}(x)\).  We compare it with the MAP
profile using
\begin{equation}
    E_{L^2}^{\rm MCMC/MAP}
    =
    \left[
    \frac{
        \displaystyle
        \int
        \left|
            \overline f^{\,\rm MCMC}_{n,\theta}(x)
            -
            f^{\rm MAP}_{n,\theta}(x)
        \right|^2
        \,\du x
    }{
        \displaystyle
        \int
        \left|
            f^{\rm MAP}_{n,\theta}(x)
        \right|^2
        \,\du x
    }
    \right]^{1/2}.
    \label{eq:quartic_mcmc_map_l2}
\end{equation}

Table~\ref{tab:quartic_mcmc_l2} gives the discrepancies along the
continuation sequence.  They decrease overall with increasing \(n\).
At \(n=8000\), the values range from \(0.0060\) to \(0.0093\), so the
sampled mean and MAP profiles agree at the percent level or better for
all four deformation strengths.

\begin{table}[t]
\caption{
Relative \(L^2\) discrepancy between the corner-transfer MCMC mean and
the MAP profile.
}
\label{tab:quartic_mcmc_l2}
\centering
\begingroup
\renewcommand{\arraystretch}{1.12}
\setlength{\tabcolsep}{7pt}
\begin{ruledtabular}
\begin{tabular}{cccccc}
\(\theta\)
&
\(n=500\)
&
\(n=1000\)
&
\(n=2000\)
&
\(n=4000\)
&
\(n=8000\)
\\
\hline
\(0.0\) & \(0.0208\) & \(0.0200\) & \(0.0160\) & \(0.0127\) & \(0.0060\) \\
\(0.1\) & \(0.0279\) & \(0.0195\) & \(0.0155\) & \(0.0120\) & \(0.0091\) \\
\(0.5\) & \(0.0298\) & \(0.0225\) & \(0.0164\) & \(0.0114\) & \(0.0093\) \\
\(1.0\) & \(0.0324\) & \(0.0238\) & \(0.0193\) & \(0.0123\) & \(0.0091\)
\end{tabular}
\end{ruledtabular}
\endgroup
\end{table}

\begin{figure*}[t]
    \centering
    \includegraphics[width=0.80\textwidth]
    {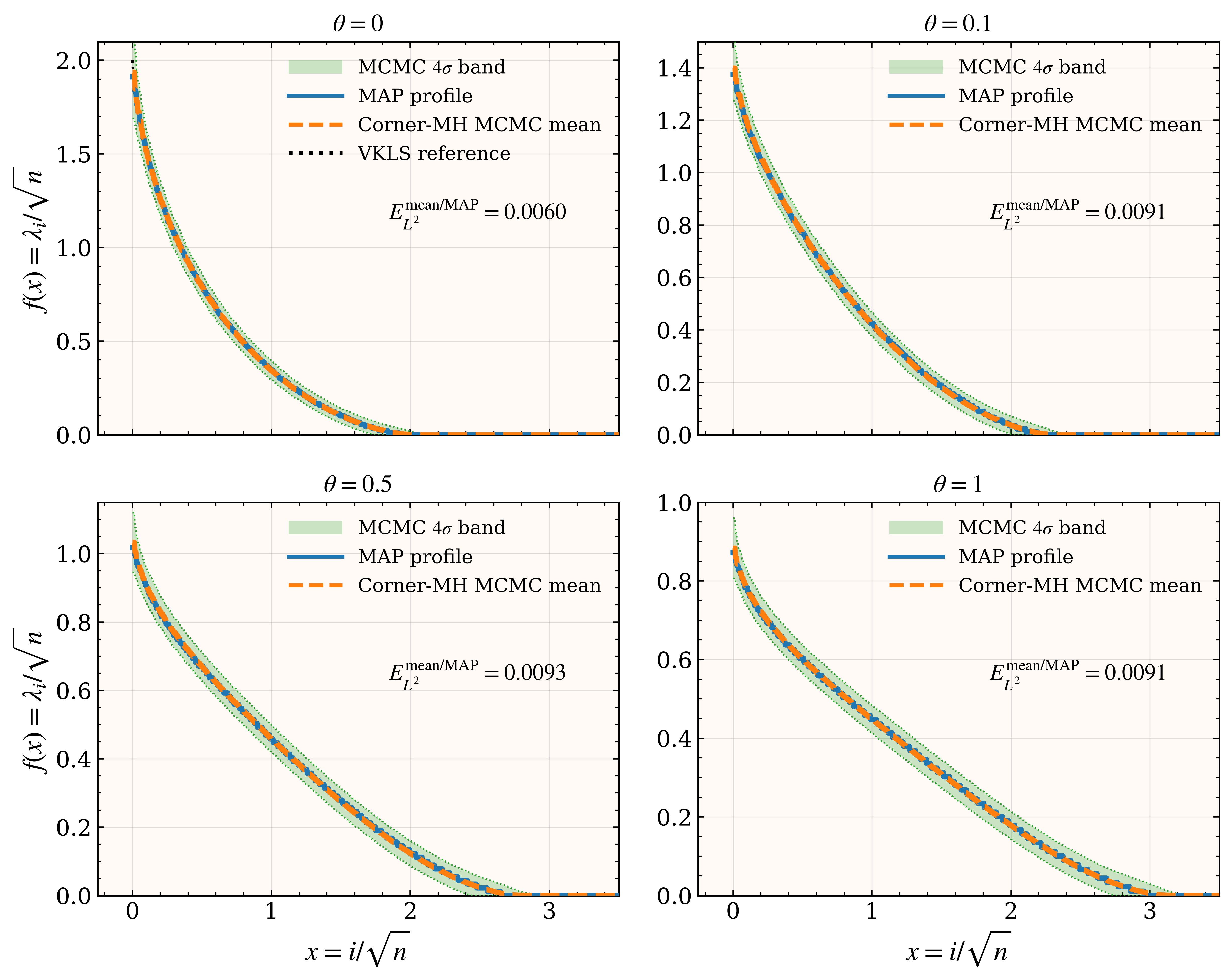}
    \caption{
    MAP and corner-transfer Metropolis--Hastings mean profiles at
    \(n=8000\) and \(T=1\).
    Solid blue curves show the MAP profiles, while dashed orange curves
    show the sampled means.  The shaded regions indicate four pointwise sample standard deviations
across the retained partitions; they are fluctuation bands rather than
uncertainty intervals for the sampled mean.  The VKLS curve is included as an
    undeformed reference in the \(\theta=0\) panel.
    }
    \label{fig:quartic_mcmc_map}
\end{figure*}

Figure~\ref{fig:quartic_mcmc_map} shows the comparison at \(n=8000\).
The sampled mean follows the MAP profile across the visible support.
Finite-temperature sampling produces a small broadening of the profile,
but no competing macroscopic shape or anomalous long tail is observed.

The edge observables give the same picture.  At \(n=8000\), the MCMC
means of both \(\lambda_1/\sqrt n\) and
\(\ell(\lambda)/\sqrt n\) differ from the corresponding MAP values by
less than approximately \(2\%\).  These shifts are small compared with
the deformation-induced changes reported in
Table~\ref{tab:quartic_deformation_observables}.

The sampled mean action also lies slightly above the MAP action, with
\begin{equation}
    3.3\times10^{-3}
    \leq
    \frac{
        \left\langle S^{(1)}_{n,\theta}\right\rangle_{\rm MCMC}
        -
        S^{(1)}_{n,\theta}
        \left(\lambda^{\rm MAP}_{n,\theta}\right)
    }{n}
    \leq
    4.2\times10^{-3}
    \qquad
    (n=8000).
    \label{eq:quartic_mcmc_action_gap}
\end{equation}
The positive gap is consistent with finite-temperature fluctuations
around a finite-size minimum.

Taken together, the discrete continuation, the large-\(n\) neural
calculation, and the \(T=1\) sampling give a consistent deformation
family.  Increasing \(\theta\) suppresses the leading rows and broadens
the support, 
while the separately computed neural, MAP, and MCMC profiles agree
at the percent level in the positively deformed cases.
These results provide numerical evidence for a common macroscopic saddle
shape at each deformation strength considered here, without assuming an
analytical deformed profile.
Acceptance rates, chain-to-chain comparisons, and trace diagnostics are
reported in Appendix~\ref{app:mcmc_details}.

\clearpage

\section{Discussion and conclusions}
\label{sec:discussion_conclusion}

The calculations considered here show that no single numerical
representation is equally natural for all of these Young-diagram
ensembles.
The appropriate variational variables depend on the
structure of the ensemble.  Hook-weighted measures retain a nonlocal
dependence on the rows and columns of the diagram, uniformly weighted
partitions are more naturally described through coarse-grained entropy
densities, and fixed-\(q\) \(q\)-Plancherel diagrams require discrete
row fractions rather than the balanced \(i/\sqrt n\) scaling.  The main
role of the neural model is therefore not to replace this structure, but
to provide a flexible parametrization within it.

The benchmark calculations support this approach.  The finite-size
soft-hook objective recovers the Plancherel profile without using the
VKLS curve during training.  The density-based calculations reproduce
the finite-grid stationary profiles of uniform and minimal-difference
partitions, with discrepancies ranging from order \(10^{-3}\) to
\(10^{-4}\).  In the fixed-\(q\) \(q\)-Plancherel ensemble, the solver
also recovers the geometric row fractions over a broad range of \(q\),
including the more extended \(q=0.95\) case.  This last example is
particularly useful because it shows that the framework is not tied to
the \(O(\sqrt n)\) scaling of balanced diagrams.  The numerical
representation must change when the asymptotic scaling changes.

The quartically deformed hook-length ensemble provides a different test,
since no analytical deformed profile is assumed.  For the \(c_h=1\)
ensemble studied here, increasing the deformation strength suppresses
the longest rows and redistributes the fixed area over a broader support.
This trend is seen in the finite-size MAP-candidate sequence, in the
neural calculation at \(n=2\times10^{7}\), and in the mean profiles
obtained from \(T=1\) corner-transfer sampling.
The neural and MAP profiles agree
at approximately the one-percent level for the positively deformed
cases, and the MCMC means approach the MAP profiles with increasing
\(n\), also reaching percent-level agreement at \(n=8000\).  The
consistency of these three calculations is the main evidence for the
deformation-dependent saddle family reported here.

The comparison also clarifies the relation between the different
numerical objects.  A finite-size MAP diagram maximizes the probability
of an individual partition, whereas a typical macroscopic profile is
selected by the concentration of probability mass over many microscopic
diagrams.  These two notions need not coincide in a general ensemble.
For the deformed hook model, the agreement between the MAP profiles and
the finite-temperature MCMC means indicates that they describe the same
macroscopic saddle region over the range of sizes considered.  This
conclusion relies on the sampling calculation and should not be inferred
from the MAP search alone.

Several limitations remain.  The results do not constitute a proof of
existence or uniqueness of a deformed limit shape, nor do they determine
a convergence rate or an analytical equation for the limiting profile.
The neural calculation supplies one very-large-\(n\) realization rather
than a full neural cross-\(n\) sequence.  The discrete MAP and MCMC
calculations reach smaller sizes, and their residual finite-size effects
are most visible near the profile edges.  In addition, the deformed model
uses \(c_h=1\).  Its undeformed MAP diagrams coincide with those of the
ordinary Plancherel hook action, but its finite-temperature measure is
not the strict \(c_h=2\) Plancherel measure.  The conclusions for the
deformed ensemble should therefore be understood within this
Plancherel-type hook model.

A natural next step is to derive the continuum variational problem
associated with the quartic deformation and compare its stationary
equation directly with the numerical profiles.  It would also be useful
to extend the discrete calculations to larger \(n\), examine the
crossover between fixed-\(q\) and near-Plancherel scaling, and study
deformations for which several competing saddle shapes may occur.  Such
problems would provide a sharper test of whether the same
structure-preserving strategy can distinguish a unique limit shape from
coexistence or metastability.

In summary, the results show that neural variational methods can recover
known Young-diagram profiles and produce consistent large-\(n\)
predictions when no analytical saddle is available, provided that the
discrete geometry, normalization, and ensemble-specific variational
structure are built into the formulation.  The quartically deformed
example illustrates the main advantage of this approach: 
the neural profile is not used as a stand-alone prediction, but is
tested against exact-action discrete optimization and direct sampling
of the underlying finite-size ensemble.

\begin{acknowledgments}
The authors acknowledge HIAS for access to the ``Quantum Universe
Physical Simulation Platform''.  Q.C. acknowledges support from the
National Natural Science Foundation of China under Grant
No.~12341101 and from Grant No.~G02X2403006.  B.-X.G. acknowledges
support from the National Natural Science Foundation of China under
Grant No.~12505066.
\end{acknowledgments}

\clearpage
\appendix
\section{Numerical details of the variational and
discrete solvers}
\label{app:method_details}

This Appendix records the implementation details omitted from
Sec.~\ref{sec:method}.  All five ensembles are treated with neural
solvers, but their finite-grid variables and objectives depend on the
structure of the underlying measure.  The Plancherel and quartically
deformed calculations use continuous row lengths, the uniform and
minimal-difference calculations use density representations, and the
fixed-\(q\) \(q\)-Plancherel calculation acts on a finite sequence of row
fractions.

The three hook-based solvers also use different row grids and occupancy
conventions.  We describe these discretizations separately rather than
introducing a common convention that was not used in the calculations.

The neural checkpoints reported in the main text are selected by the
corresponding variational evaluation objectives.  Analytical profiles
are used only in post-optimization comparisons.  The MAP partitions and
MCMC mean profiles likewise do not enter the training or checkpoint
selection of the quartic neural solver.

\subsection{Network architecture and numerical settings}
\label{app:finite_grid_neural}

All neural calculations use fully connected coordinate networks with five
hidden layers of width \(128\) and SiLU activation functions.  Linear
weights are initialized with Xavier initialization and biases are set to
zero.  The reported calculations use single-precision floating-point
arithmetic and random seed \(1234\).  Optimization is performed with
AdamW, a cosine-annealed learning rate, and gradient clipping.

Table~\ref{tab:appendix_neural_settings} summarizes the principal
settings used for the figures and numerical comparisons in the main
text.  Here \(M\) denotes the number of grid points or retained rows.

\begin{table}[t]
\caption{
Principal settings of the neural calculations reported in the main
text.  Detailed learning rates and stage lengths are given in the
ensemble-specific subsections below.
}
\label{tab:appendix_neural_settings}
\centering
\begingroup
\small
\renewcommand{\arraystretch}{1.35}
\setlength{\tabcolsep}{4pt}

\begin{tabular}{
|p{0.20\textwidth}
 p{0.23\textwidth}
 p{0.20\textwidth}
 p{0.27\textwidth}|
}
\hline

\textbf{Ensemble}
&
\textbf{Parameters and size}
&
\textbf{Grid}
&
\textbf{Optimization}
\\
\hline

\textbf{Plancherel}
&
\(n=10^{5}\)
&
\(\begin{gathered}
X_{\max}=2.05,\\
M=649
\end{gathered}\)
&
Two stages:
\(\tau=0.8,\,0.4\)
\\
\hline

\textbf{Uniform partitions}
&
\(n=10^{5}\)
&
\(\begin{gathered}
X_{\max}=10,\\
K=3163
\end{gathered}\)
&
Single \(4000\)-epoch run
\\
\hline

\textbf{Minimal-difference partitions}
&
\(\begin{gathered}
p=1,2,3,\\
n=10^{4}
\end{gathered}\)
&
\(\begin{gathered}
X_{\max}=10,\\
K=1000
\end{gathered}\)
&
Two \(2500\)-epoch stages
\\
\hline

\textbf{\(q\)-Plancherel}
&
\(\begin{gathered}
q=0.2,0.4,0.6,0.8,0.95,\\
n=5\times10^{4}
\end{gathered}\)
&
\(M=2237\)
&
Single \(3000\)-epoch run;
\(\tau:1\rightarrow0.12\)
\\
\hline

\textbf{Quartic deformation}
&
\(\begin{gathered}
\theta=0,0.1,0.5,1,\\
n=2\times10^{7}
\end{gathered}\)
&
\(\begin{gathered}
X_{\max}=3.6,\\
M=16100
\end{gathered}\)
&
Four stages:
\(\tau=0.8,\,0.4,\,0.2,\,0.1\)
\\
\hline

\end{tabular}
\endgroup
\end{table}

The Plancherel and quartic calculations represent continuous row lengths
directly.  The uniform and minimal-difference calculations construct the
density \(\rho=-f'\) and recover the Young-diagram boundary through a
discrete tail sum.  The \(q\)-Plancherel calculation uses a finite-row
representation adapted to row lengths of order \(n\).

\subsection{Hook-action neural solvers}
\label{app:hook_neural_details}

\subsubsection{Ordinary Plancherel calculation}
\label{app:plancherel_neural_details}

For the ordinary Plancherel benchmark, the retained row coordinates are
\begin{equation}
    x_i=\frac{i}{\sqrt n},
    \qquad
    i=1,\ldots,M,
    \qquad
    M=\left\lceil X_{\max}\sqrt n\right\rceil .
    \label{eq:app_plancherel_grid}
\end{equation}
The reported calculation uses \(n=10^{5}\), \(X_{\max}=2.05\), and
\(M=649\).  The coordinate supplied to the network is linearly mapped to
\([-1,1]\).

The network output is converted to monotone row lengths through
\begin{equation}
    q_i
    =
    {\rm softplus}\!\left(N_\phi(x_i)\right)+\epsilon,
    \qquad
    T_i=\sum_{k=i}^{M}q_k,
    \qquad
    \lambda_i^{(\phi)}
    =
    n\,
    \frac{T_i}{\displaystyle\sum_{r=1}^{M}T_r},
    \label{eq:app_plancherel_rows}
\end{equation}
with \(\epsilon=10^{-8}\).  This construction ensures that the rows are
nonnegative and nonincreasing and that
\(\sum_i\lambda_i^{(\phi)}=n\).

The softened occupancy is
\begin{equation}
    O_{ij}^{(\phi)}
    =
    \sigma\left(
        \frac{\lambda_i^{(\phi)}-j}{\tau}
    \right).
    \label{eq:app_plancherel_occupancy}
\end{equation}
The number of columns is adjusted during training according to
\begin{equation}
    J
    =
    \min\left\{
        n,\,
        \max\left[
            M,\,
            \left\lceil\max_i\lambda_i^{(\phi)}\right\rceil
            +
            \max\!\left(10,\left\lceil8\tau\right\rceil\right)
        \right]
    \right\}.
    \label{eq:app_plancherel_column_cutoff}
\end{equation}
This prevents a long first row from extending beyond the column range
included in the hook calculation.

The soft column lengths and hook lengths are
\begin{equation}
    \lambda_j^{\prime(\phi)}
    =
    \sum_{i=1}^{M}O_{ij}^{(\phi)},
    \qquad
    h_{ij}^{(\phi)}
    =
    {\rm softplus}_{\beta=3}
    \left(
        \lambda_i^{(\phi)}
        -j
        +
        \lambda_j^{\prime(\phi)}
        -i
        +1
    \right)
    +
    \epsilon .
    \label{eq:app_plancherel_soft_hooks}
\end{equation}
The training objective is
\begin{equation}
    \mathcal{L}_{\rm Pl}^{(n)}
    =
    {}
    \frac{1}{n}
    \sum_{i=1}^{M}
    \sum_{j=1}^{J}
    O_{ij}^{(\phi)}
    \log h_{ij}^{(\phi)}
    +
    w_A
    \left[
        \frac{1}{n}
        \sum_{i=1}^{M}
        \sum_{j=1}^{J}
        O_{ij}^{(\phi)}
        -1
    \right]^2,
    \label{eq:app_plancherel_loss}
\end{equation}
where \(w_A=10^{-2}\).  The second term corrects the small difference
between the exact row normalization and the area represented by the
finite-temperature occupancy field.  Column sums are evaluated in blocks
of \(512\) to limit memory use.

Training proceeds through two stages, with
\(\tau=0.8\) and \(0.4\).  The stages contain at most \(1500\) and
\(2500\) epochs, with initial learning rates \(10^{-4}\) and
\(5\times10^{-5}\), respectively.  Continuation between stages and the
reported final profile use the checkpoint with the smallest soft-hook
evaluation loss.  The VKLS profile is recorded only as an external
diagnostic and is not used for training or continuation.

\subsubsection{\(q\)-Plancherel calculation}
\label{app:qplancherel_neural_details}

The fixed-\(q\) regime is not represented on the balanced
\(i/\sqrt n\) grid.  Instead, the solver retains
\begin{equation}
    M=\max\left\{240,\left\lceil10\sqrt n\right\rceil\right\}
\end{equation}
candidate rows and constructs a finite sequence of continuous row
lengths.  For the reported calculations, \(n=5\times10^{4}\), giving
\(M=2237\).

For row \(i\), the network input is
\begin{equation}
    \left(
        \frac{i}{M},
        \frac{\log(1+i)}{\log(1+M)},
        q,
        \frac{\log(1-q)}{10},
        \frac{\log n}{10}
    \right).
    \label{eq:app_qplancherel_features}
\end{equation}
Its output defines positive row decrements,
\begin{equation}
    a_i={\rm softplus}\!\left(N_\phi(i,q,n)\right)+10^{-10}.
\end{equation}
The continuous rows are obtained from
\begin{equation}
    \widetilde{\lambda}_i=\sum_{k=i}^{M}a_k,
    \qquad
    \lambda_i^{(\phi)}
    =
    n\,
    \frac{\widetilde{\lambda}_i}
    {\displaystyle\sum_{r=1}^{M}\widetilde{\lambda}_r}.
    \label{eq:app_qplancherel_rows}
\end{equation}
They are therefore nonnegative and nonincreasing, with
\(\sum_i\lambda_i^{(\phi)}=n\).

The softened occupancy uses the cell-centered convention
\begin{equation}
    O_{ij}^{(\phi)}
    =
    \sigma\left(
        \frac{\lambda_i^{(\phi)}-j+\frac12}{\tau}
    \right).
    \label{eq:app_qplancherel_occupancy}
\end{equation}
All columns \(j=1,\ldots,n\) are retained.  This prevents the network
from placing boxes beyond the column range included in the hook
objective.

For numerical stability, the logarithm of the \(q\)-integer is evaluated
as
\begin{equation}
    \log[h]_q
    =
    \log\left(1-q^h\right)-\log(1-q),
\end{equation}
using an implementation based on \texttt{expm1} for \(q<1\), while the
\(q\to1\) limit is evaluated as \(\log h\).

The neural objective is
\begin{equation}
    \mathcal{L}_{q{\rm Pl}}^{(n)}
    =
    \frac{1}{n}
    \left[
        \sum_{i,j}
        O_{ij}^{(\phi)}
        \left(
            \log h_{ij}^{(\phi)}
            +
            \log[h_{ij}^{(\phi)}]_q
        \right)
        -
        \log q
        \sum_i(i-1)\lambda_i^{(\phi)}
    \right]
    +
    w_{\rm sm}\mathcal{R}_{\rm sm}
    +
    w_{\rm tail}\mathcal{R}_{\rm tail}.
    \label{eq:app_qplancherel_loss}
\end{equation}
The two weak stabilizers are
\begin{equation}
    \mathcal{R}_{\rm sm}
    =
    \frac{1}{M-1}
    \sum_{i=1}^{M-1}
    \left[
        \log(a_{i+1}+\epsilon)
        -
        \log(a_i+\epsilon)
    \right]^2,
    \qquad
    \mathcal{R}_{\rm tail}
    =
    \left(
        \frac{\lambda_M^{(\phi)}}{n}
    \right)^2,
\end{equation}
with \(w_{\rm sm}=10^{-6}\) and \(w_{\rm tail}=10^{-7}\).  These terms
regularize the finite-row representation and contain no analytical
profile information.

Separate runs are performed for
\(q=0.2,0.4,0.6,0.8,\) and \(0.95\).  Each run uses \(3000\) epochs,
with \(\tau\) reduced geometrically from \(1\) to \(0.12\) and the
learning rate cosine annealed from \(2\times10^{-3}\) to
\(2\times10^{-4}\).  Checkpoints are selected by evaluating the same
objective at the target value of \(q\) and the final temperature
\(\tau=0.12\).  The \(q\)-homotopy, L-BFGS polishing, and post-training
hard-action local search are disabled in the reported calculations.

\subsubsection{Quartically deformed hook-length ensemble}
\label{app:quartic_neural_details}

The neural quartic calculation uses the same finite-size action as the
discrete MAP and MCMC calculations,
\begin{equation}
    S^{(1)}_{n,\theta}(\lambda)
    =
    \sum_{u\in\lambda}\log h(u)
    +
    \theta\sqrt n
    \sum_i
    \left(
        \frac{\lambda_i}{\sqrt n}
    \right)^4.
    \label{eq:app_quartic_action}
\end{equation}
The neural optimization is nevertheless separate from the discrete
calculations.  MAP partitions and MCMC profiles do not enter the loss,
initialization, or checkpoint selection.

The reported calculations use \(n=2\times10^{7}\) and
\(X_{\max}=3.6\).  The cell-centred row grid is
\begin{equation}
    x_i
    =
    \frac{i-\frac12}{\sqrt n},
    \qquad
    i=1,\ldots,M,
    \qquad
    M=\left\lceil X_{\max}\sqrt n\right\rceil=16100.
    \label{eq:app_quartic_grid}
\end{equation}
The coordinate supplied to the network is linearly mapped to
\([-1,1]\).

The network produces a positive generator
\begin{equation}
    q_i
    =
    {\rm softplus}\!\left(N_\phi(x_i)\right)
    +
    \epsilon,
    \qquad
    \epsilon=10^{-8}.
\end{equation}
Its right-to-left trapezoidal tail is
\begin{equation}
    T_i
    =
    \sum_{k=i}^{M-1}
    \frac{x_{k+1}-x_k}{2}
    \left(q_k+q_{k+1}\right),
    \qquad
    T_M=0,
    \label{eq:app_quartic_tail}
\end{equation}
and the continuous row lengths are
\begin{equation}
    \lambda_i^{(\phi)}
    =
    n\,
    \frac{T_i}
    {\displaystyle\sum_{r=1}^{M}T_r}.
    \label{eq:app_quartic_rows}
\end{equation}
The rows are therefore nonnegative and nonincreasing, with
\(\sum_i\lambda_i^{(\phi)}=n\).  The terminal condition \(T_M=0\)
matches the finite-support condition in the continuum tail-integral
representation.

The softened occupancy uses the cell-centred convention
\begin{equation}
    O_{ij}^{(\phi)}
    =
    \sigma\left(
        \frac{\lambda_i^{(\phi)}-j+\frac12}{\tau}
    \right),
    \label{eq:app_quartic_occupancy}
\end{equation}
so that the transition lies at the boundary of the \(j\)th unit cell.
The corresponding soft column lengths and hook lengths are
\begin{equation}
    \lambda_j^{\prime(\phi)}
    =
    \sum_{i=1}^{M}O_{ij}^{(\phi)},
    \qquad
    h_{ij}^{(\phi)}
    =
    {\rm softplus}_{\beta=3}
    \left(
        \lambda_i^{(\phi)}
        -j
        +
        \lambda_j^{\prime(\phi)}
        -i
        +1
    \right)
    +
    \epsilon.
    \label{eq:app_quartic_soft_hooks}
\end{equation}

The number of columns is adjusted during optimization according to
\begin{equation}
    J
    =
    \min\left\{
        n,\,
        \max\left[
            M,\,
            \left\lceil\max_i\lambda_i^{(\phi)}\right\rceil
            +
            \max\left(10,\left\lceil8\tau\right\rceil\right)
        \right]
    \right\}.
    \label{eq:app_quartic_column_cutoff}
\end{equation}
Column sums are evaluated in blocks of \(512\) to control GPU memory
usage.

For convenience, define the soft-occupancy area and the terminal
stabilizer by
\begin{equation}
    A_\tau
    =
    \frac{1}{n}
    \sum_{i=1}^{M}
    \sum_{j=1}^{J}
    O_{ij}^{(\phi)},
    \qquad
    \mathcal{R}_T
    =
    \frac{1}{N_T}
    \sum_{i=M-N_T+1}^{M}
    \left(
        \frac{\lambda_i^{(\phi)}}{\sqrt n}
    \right)^2.
    \label{eq:app_quartic_auxiliary_terms}
\end{equation}
The training objective is then
\begin{equation}
    \mathcal{L}^{(n)}_{\rm quartic}
    =
    \frac{1}{n}
    \sum_{i=1}^{M}
    \sum_{j=1}^{J}
    O_{ij}^{(\phi)}
    \log h_{ij}^{(\phi)}
    +
    \frac{\theta}{\sqrt n}
    \sum_{i=1}^{M}
    \left(
        \frac{\lambda_i^{(\phi)}}{\sqrt n}
    \right)^4
    +
    w_A(A_\tau-1)^2
    +
    w_T\mathcal{R}_T.
    \label{eq:app_quartic_full_loss}
\end{equation}
The reported runs use
\(w_A=10^{-1}\), \(w_T=10^{-3}\), and \(N_T=12\).
The area term corrects the difference between the exactly normalized
continuous rows and the area represented by the finite-temperature
occupancy field.  The terminal term weakly suppresses leakage at the
finite computational boundary and contains no reference-profile
information.

Training proceeds through the four stages listed in
Table~\ref{tab:appendix_quartic_schedule}.  At each stage the learning
rate follows a cosine schedule, and the checkpoint with the lowest
soft-action evaluation loss is passed to the next stage.

\begin{table}[t]
\caption{
Temperature continuation used in the neural quartic calculation.
The epoch count is the maximum allowed before early stopping.
}
\label{tab:appendix_quartic_schedule}
\centering
\begin{ruledtabular}
\begin{tabular}{cccc}
Stage &
\(\tau\) &
Maximum epochs &
\(\eta_{\rm start}\rightarrow\eta_{\rm end}\)
\\
1 & \(0.8\) & \(1000\) & \(10^{-4}\rightarrow2\times10^{-5}\) \\
2 & \(0.4\) & \(2000\) & \(5\times10^{-5}\rightarrow10^{-5}\) \\
3 & \(0.2\) & \(4000\) & \(2.5\times10^{-5}\rightarrow5\times10^{-6}\) \\
4 & \(0.1\) & \(8000\) & \(10^{-5}\rightarrow2\times10^{-6}\)
\end{tabular}
\end{ruledtabular}
\end{table}

The early-stopping patience values are \(500\), \(700\), \(900\), and
\(1000\) epochs, respectively, with a minimum improvement of \(10^{-7}\).
Gradient norms are clipped at \(10\), and the AdamW weight decay is
\(10^{-6}\).

The four deformation strengths
\(\theta=0,0.1,0.5,\) and \(1\) are trained sequentially.  The best
network at one value initializes the next, following
\(0\rightarrow0.1\rightarrow0.5\rightarrow1\).  Only the network
parameters are transferred; no MAP partition, MCMC sample, or analytical
profile is used in this continuation.

The code also permits an exact local refinement of the projected neural
partition.  This option is disabled in all results reported here.  The
neural--MAP comparison therefore involves the unrefined neural output and
an independently obtained MAP candidate.

\subsection{Entropy-based neural solvers}
\label{app:entropy_neural_details}

\subsubsection{Uniform random partitions}
\label{app:uniform_neural_details}

For uniform random partitions, we use the grid
\begin{equation}
    x_k=\frac{k}{\sqrt n},
    \qquad
    \Delta x=\frac{1}{\sqrt n},
    \qquad
    k=1,\ldots,K,
    \qquad
    K=\left\lceil X_{\max}\sqrt n\right\rceil .
    \label{eq:app_uniform_grid}
\end{equation}
The reported calculation uses \(n=10^{5}\) and \(X_{\max}=10\), giving
\(K=3163\).  The grid starts at \(x_1=1/\sqrt n\), since the network input
contains \(\log x\) and the continuum profile is singular at \(x=0\).

The two input features \(x\) and \(\log x\) are mapped separately to
\([-1,1]\) over the computational domain.  The network produces
\begin{equation}
    q_k
    =
    {\rm softplus}\!\left(N_\phi(x_k,\log x_k)\right)
    +
    \epsilon .
\end{equation}
The density is normalized through
\begin{equation}
    A_\phi
    =
    \Delta x
    \sum_{k=1}^{K}x_kq_k,
    \qquad
    \rho_k
    =
    \frac{q_k}{A_\phi+\epsilon},
    \qquad
    f_k
    =
    \Delta x
    \sum_{\ell=k}^{K}\rho_\ell .
    \label{eq:app_uniform_density_profile}
\end{equation}
The tail sum makes \(f_k\) nonnegative and nonincreasing, while the
normalization enforces the discrete area condition up to the numerical
offset \(\epsilon\).

The implemented objective is
\begin{equation}
    \mathcal{L}_{\rm unif}
    =
    -
    \Delta x
    \sum_{k=1}^{K}
    \left[
        (1+\rho_k)\log(1+\rho_k)
        -
        \rho_k\log\rho_k
    \right]
    +
    w_{\rm tail}\rho_K^2
    +
    \frac{w_{\rm sm}}{K-1}
    \sum_{k=1}^{K-1}
    \left[
        \frac{
            \log(\rho_{k+1}+\epsilon)
            -
            \log(\rho_k+\epsilon)
        }{\Delta x}
    \right]^2 .
    \label{eq:app_uniform_loss}
\end{equation}
We use \(w_{\rm tail}=w_{\rm sm}=10^{-7}\).  These weak terms control
density remaining at the finite cutoff and rapid grid-scale variation in
\(\log\rho\); neither contains the known uniform-partition profile.

Training is performed for at most \(4000\) epochs.  The learning rate is
cosine annealed from \(5\times10^{-4}\) to \(2\times10^{-5}\), with
AdamW weight decay \(10^{-7}\), gradient clipping at \(10\), and an
early-stopping patience of \(1200\) epochs.

For post-training comparison, we construct a grid-consistent stationary
reference using the Bose density on the same finite grid as the neural
calculation. The reference density is
\begin{equation}
    \rho_{k}^{\rm grid}
    =
    \frac{1}{
        \exp(\beta_n x_k)-1
    },
    \label{eq:app_uniform_grid_density}
\end{equation}
where \(\beta_n\) is determined from
\begin{equation}
    \Delta x
    \sum_{k=1}^{K}
    x_k\rho_k^{\rm grid}
    =
    1.
    \label{eq:app_uniform_grid_area}
\end{equation}
The corresponding boundary is reconstructed using the same discrete
tail sum as in the neural calculation,
\begin{equation}
    f_k^{\rm grid}
    =
    \Delta x
    \sum_{\ell=k}^{K}
    \rho_\ell^{\rm grid}.
    \label{eq:app_uniform_grid_profile}
\end{equation}
For the grid used in the main calculation,
\begin{equation}
    \beta_n=1.2817451,
    \qquad
    \frac{\pi}{\sqrt6}=1.2825498.
\end{equation}

The finite-grid reference is distinct from the continuum profile
\begin{equation}
    f_{\rm cont}(x)
    =
    -\frac{\sqrt6}{\pi}
    \log\left[
        1-\exp\left(
            -\frac{\pi x}{\sqrt6}
        \right)
    \right].
    \label{eq:app_uniform_continuum_profile}
\end{equation}
The former uses the same lower cutoff, area constraint, and tail-sum
reconstruction as the numerical optimization.  It therefore isolates
the neural optimization error, whereas comparison with
\(f_{\rm cont}\) also includes finite-grid and endpoint effects.  Both
references are evaluated only after training.

\subsubsection{Minimal-difference partitions}
\label{app:minimal_difference_neural_details}

The reported minimal-difference calculation uses
\(p=1,2,3\), \(n=10^{4}\), and \(X_{\max}=10\).  The grid is
\begin{equation}
    x_k=\frac{k}{\sqrt n},
    \qquad
    \Delta x=\frac{1}{\sqrt n},
    \qquad
    k=1,\ldots,K,
    \qquad
    K=\left\lceil X_{\max}\sqrt n\right\rceil=1000.
    \label{eq:app_mdp_grid}
\end{equation}
A single parameter-conditioned network is trained simultaneously on the
three values of \(p\).

For each \(p\), we introduce the Euler--Lagrange coordinate
\begin{equation}
    z=\beta_p x .
\end{equation}
The network inputs are
\begin{equation}
    \left(
        \frac{2z}{z_{\rm sc}}-1,\,
        \frac{2\log(1+z)}{\log(1+z_{\rm sc})}-1,\,
        \frac{\log(1+p)}{\log(1+p_{\rm sc})}
    \right),
    \label{eq:app_mdp_features}
\end{equation}
with \(z_{\rm sc}=10\) and \(p_{\rm sc}=5\).

The network supplies a bounded correction
\begin{equation}
    C_\phi(z,p)
    =
    C_{\max}\tanh N_\phi(z,p),
    \qquad
    C_{\max}=2,
\end{equation}
which is tapered at large \(z\) through
\begin{equation}
    T(z)
    =
    \sigma\left(
        \frac{z_0-z}{w_z}
    \right),
    \qquad
    z_0=4.5,
    \qquad
    w_z=0.8.
\end{equation}
The Euler--Lagrange variable and density are then
\begin{equation}
    u_\phi(x;p)
    =
    \beta_p x
    -
    T(\beta_p x)C_\phi(\beta_p x,p),
    \qquad
    \rho_\phi(x;p)
    =
    \left(s_p'\right)^{-1}
    \!\left(u_\phi(x;p)\right).
    \label{eq:app_mdp_density}
\end{equation}

For \(p=1\) and \(p=2\), the inverse maps are evaluated analytically:
\begin{equation}
    \rho(u;1)
    =
    \frac{1}{e^u+1},
    \qquad
    \rho(u;2)
    =
    \frac12
    \left(
        1-
        \frac{1}{\sqrt{1+4e^{-u}}}
    \right).
    \label{eq:app_mdp_closed_inverse}
\end{equation}
For \(p\geq3\), the inverse is found by bisection over
\(0<\rho<1/p\).  The \(p=3\) calculation uses \(70\) bisection
iterations.

At each forward evaluation, \(\beta_p\) is determined from the discrete
area condition
\begin{equation}
    \Delta x
    \sum_{k=1}^{K}
    x_k\rho_\phi(x_k;p)
    =
    1.
    \label{eq:app_mdp_area}
\end{equation}
The bisection begins with
\(\beta_{\rm lo}=10^{-5}\) and \(\beta_{\rm hi}=20\), enlarging the
upper endpoint when necessary, and uses \(44\) iterations.  Gradients are
not propagated through this projection.

The training objective is
\begin{equation}
    \mathcal{L}_{p}
    =
    -
    \Delta x
    \sum_{k=1}^{K}
    s_p\!\left(\rho_\phi(x_k;p)\right)
    +
    \beta_p
    \Delta x
    \sum_{k=1}^{K}
    x_k\rho_\phi(x_k;p)
    +
    w_C
    \left\langle
        \left[T(z)C_\phi(z,p)\right]^2
    \right\rangle
    +
    w_{\rm sm}
    \left\langle
        \left[
            \frac{\Delta(TC_\phi)}{\Delta z}
        \right]^2
    \right\rangle .
    \label{eq:app_mdp_loss}
\end{equation}
Here the angle brackets denote averages over the finite grid.  Although
the projected area is unity, the term proportional to \(\beta_p\)
supplies the constrained local gradient when \(\beta_p\) is held fixed
during differentiation.

Training consists of two stages of at most \(2500\) epochs each.  The
first uses \(w_C=2\times10^{-3}\), \(w_{\rm sm}=10^{-6}\), and a learning
rate decreasing from \(5\times10^{-4}\) to \(5\times10^{-5}\).  The
second uses \(w_C=10^{-2}\), \(w_{\rm sm}=10^{-6}\), and a learning rate
decreasing from \(10^{-4}\) to \(10^{-5}\).  The weak correction terms
keep the density close to the Euler--Lagrange backbone and contain no
analytical reference-profile information.

For post-training validation, we construct a grid-consistent stationary
reference independently of the neural calculation.  For each \(p\), the
reference density is
\begin{equation}
    \rho_{p,k}^{\rm grid}
    =
    \left(s_p'\right)^{-1}
    \left(
        \beta_{p,{\rm grid}}x_k
    \right),
    \label{eq:app_mdp_grid_reference_density}
\end{equation}
where \(\beta_{p,{\rm grid}}\) is fixed by
\begin{equation}
    \Delta x
    \sum_{k=1}^{K}
    x_k\rho_{p,k}^{\rm grid}
    =
    1.
    \label{eq:app_mdp_grid_reference_area}
\end{equation}
The corresponding Young-diagram boundary is reconstructed with the same
tail-sum convention as the neural profile,
\begin{equation}
    f_{p,k}^{\rm grid}
    =
    \Delta x
    \sum_{\ell=k}^{K}
    \rho_{p,\ell}^{\rm grid}.
    \label{eq:app_mdp_grid_reference_profile}
\end{equation}
This reference uses the same finite grid, lower and upper cutoffs,
discrete area normalization, and boundary reconstruction as the neural
calculation.  It therefore largely separates the neural approximation
error from the finite-grid effects common to both profiles.

The area multipliers obtained from the neural calculation and from the
independently constructed grid reference are
\begin{equation}
\begin{array}{c|cc}
    p
    &
    \beta_{p,{\rm NN}}
    &
    \beta_{p,{\rm grid}}
    \\
    \hline
    1 & 0.9056638 & 0.9062571 \\
    2 & 0.8089981 & 0.8094489 \\
    3 & 0.7491057 & 0.7494876
\end{array}.
\label{eq:app_mdp_beta_comparison}
\end{equation}
The discrete areas of the reported neural profiles differ from unity by
less than \(1.2\times10^{-7}\).

For completeness, the full-domain absolute \(L^2\) discrepancies from
the grid references are
\begin{equation}
\begin{split}
    E_{L^2}^{1,{\rm grid},{\rm full}}
    &=
    3.2349\times10^{-4},
    \\
    E_{L^2}^{2,{\rm grid},{\rm full}}
    &=
    2.2296\times10^{-4},
    \\
    E_{L^2}^{3,{\rm grid},{\rm full}}
    &=
    1.7936\times10^{-4}.
\end{split}
\label{eq:app_mdp_full_l2}
\end{equation}
These values are close to the cut-domain discrepancies reported in the
main text.  The agreement is therefore not produced by removing the
first few grid cells.  The reference profiles are evaluated only after
training and do not enter the loss or checkpoint selection.

\subsection{Projection to integer partitions and hard actions}
\label{app:projection_details}

The hook-action solvers produce continuous, nonincreasing row lengths.
Before evaluating the exact finite-size action, these rows are projected
onto an integer partition of \(n\).

We first define
\begin{equation}
    \widetilde{\lambda}_i
    =
    \left\lfloor\lambda_i^{(\phi)}\right\rfloor,
    \qquad
    r_i
    =
    \lambda_i^{(\phi)}
    -
    \left\lfloor\lambda_i^{(\phi)}\right\rfloor .
    \label{eq:app_projection_floor}
\end{equation}
Any small numerical violation of monotonicity is removed before the box
count is corrected.  If the integerized rows contain fewer than \(n\)
boxes, boxes are added only where the partition condition remains
satisfied.  For \(i>1\), this requires
\begin{equation}
    \widetilde{\lambda}_i+1
    \leq
    \widetilde{\lambda}_{i-1}.
\end{equation}
In the \(q\)-Plancherel and quartic implementations, admissible rows are
considered in decreasing order of \(r_i\).  For numerical robustness, an
analogous removal step is included if an excess box count occurs; a box
may be removed from row \(i\) only when
\begin{equation}
    \widetilde{\lambda}_i-1
    \geq
    \widetilde{\lambda}_{i+1},
\end{equation}
with \(\widetilde{\lambda}_{i+1}=0\) at the final row.  Rows with smaller
fractional parts are considered first.  The ordinary Plancherel
diagnostic uses the same admissibility conditions with a simpler greedy
ordering.

The resulting partition satisfies
\begin{equation}
    \lambda_1^{\rm proj}
    \geq
    \lambda_2^{\rm proj}
    \geq\cdots\geq0,
    \qquad
    \lambda_i^{\rm proj}\in\mathbb{Z}_{\geq0},
    \qquad
    \sum_i\lambda_i^{\rm proj}=n.
    \label{eq:app_projected_partition}
\end{equation}

For an integer partition, the column lengths and hook lengths are
recomputed directly:
\begin{equation}
    \lambda'_j
    =
    \#\{i:\lambda_i\geq j\},
    \qquad
    h(i,j)
    =
    \lambda_i-j+\lambda'_j-i+1.
    \label{eq:app_exact_hooks}
\end{equation}
We denote the exact hook sum by
\begin{equation}
    H_{\rm hook}(\lambda)
    =
    \sum_{(i,j)\in\lambda}\log h(i,j).
    \label{eq:app_exact_hook_sum}
\end{equation}

For \(q\)-Plancherel, the hard action is
\begin{equation}
    S_q(\lambda)
    =
    \sum_{(i,j)\in\lambda}
    \left[
        \log h(i,j)
        +
        \log[h(i,j)]_q
    \right]
    -
    \log q
    \sum_i(i-1)\lambda_i,
    \label{eq:app_hard_q_action}
\end{equation}
while the quartic action is
\begin{equation}
    S^{(1)}_{n,\theta}(\lambda)
    =
    H_{\rm hook}(\lambda)
    +
    \theta\sqrt n
    \sum_i
    \left(
        \frac{\lambda_i}{\sqrt n}
    \right)^4.
    \label{eq:app_hard_quartic_action}
\end{equation}

These exact actions are evaluated only after neural optimization.  They
do not contribute gradients and are not used for checkpoint selection.
For the quartic results reported in the main text, no subsequent
hard-action refinement is applied to the projected neural partition.

\subsection{Discrete MAP search and cross-\(n\) continuation}
\label{app:map_details}

The discrete calculation searches for low-action integer partitions using
the exact quartic action in
Eq.~\eqref{eq:app_hard_quartic_action}.  Each trial move transfers one
box while preserving the total size of the partition.

The source row is selected from a mixture of two proposals.  With
probability \(0.55\), row \(i\) is chosen with probability
\begin{equation}
    p_{\rm src}(i)=\frac{\lambda_i}{n},
\end{equation}
while the remaining proposals choose uniformly among the nonzero rows.
After one box is removed, it forms a new row with probability \(0.18\).
Otherwise, it is added to an existing row.  The destination row is chosen
with probability proportional to \(1/\sqrt{\lambda_i+1}\) in \(65\%\) of
these moves and uniformly in the remaining \(35\%\).  The resulting row
lengths are then sorted into nonincreasing order.

This proposal is used for optimization rather than equilibrium sampling,
and no Hastings correction is applied.  If
\(\Delta S=S(\lambda')-S(\lambda)\), the candidate is accepted whenever
\(\Delta S\leq0\), and otherwise with probability
\begin{equation}
    \exp\left(-\frac{\Delta S}{T}\right).
\end{equation}
The temperature decreases geometrically according to
\begin{equation}
    T(s)
    =
    T_0
    \left(
        \frac{T_{\rm final}}{T_0}
    \right)^{s/N_{\rm step}},
\end{equation}
with \(T_0=2\), \(T_{\rm final}=0.01\), and
\(N_{\rm step}=5\times10^{5}\).

At each \((n,\theta)\), the standard initial-state pool contains a
VKLS-based partition, a rectangular partition, an exponential partition,
and \(16\) random partitions.  For \(n>500\), the lowest-action partition
from the preceding size is also rescaled to the new value of \(n\), and
two independently perturbed copies are added to the pool.  Each perturbed
copy is obtained by applying approximately \(0.02n\) one-box moves to the
continued partition.

The reported calculations use
\(n=500,1000,2000,4000,\) and \(8000\).  To continue a partition from
\(n_{\rm old}\) to \(n_{\rm new}\), its scaled English profile is sampled
on the new cell-centred row grid, rounded to integer row lengths, and
corrected to contain exactly \(n_{\rm new}\) boxes.  Continuation changes
only the initial state; the objective at the new size is always the exact
action for \(n_{\rm new}\).

All initial states at fixed \((n,\theta)\) are optimized independently,
and the partition with the smallest exact action is retained as the MAP
candidate.  Although the implementation also permits continuation between
different values of \(\theta\) at fixed \(n\), that option does not enter
the reported discrete calculations, which were run separately for each
deformation strength.

\subsection{Corner-transfer Metropolis--Hastings sampling}
\label{app:mcmc_details}

The MCMC calculation samples the finite-size quartic ensemble
\begin{equation}
    \mathbb{P}_{n,\theta}(\lambda)
    \propto
    \exp\left[-S^{(1)}_{n,\theta}(\lambda)\right]
\end{equation}
at fixed temperature \(T=1\).

A row \(i\) contains a removable corner when
\begin{equation}
    \lambda_i>\lambda_{i+1},
\end{equation}
where the row below the final nonzero row is assigned length zero.  The
set of removable corners is denoted by \(R(\lambda)\).  A proposal first
selects one element of \(R(\lambda)\) uniformly and removes the
corresponding box, producing an intermediate partition \(\nu\) of size
\(n-1\).  The box is then placed uniformly at one of the addable corners
\(A(\nu)\), giving a candidate partition \(\lambda'\).

Because the initial and proposed partitions can have different numbers
of removable corners, the proposal is not symmetric.  For a move passing
through the same intermediate partition \(\nu\),
\begin{equation}
    Q(\lambda\rightarrow\lambda')
    =
    \frac{1}{|R(\lambda)|\,|A(\nu)|},
    \qquad
    Q(\lambda'\rightarrow\lambda)
    =
    \frac{1}{|R(\lambda')|\,|A(\nu)|}.
\end{equation}
The addable-corner factors cancel in the proposal ratio, and the
Metropolis--Hastings acceptance probability is
\begin{equation}
    p_{\rm acc}(\lambda\rightarrow\lambda')
    =
    \min\left\{
        1,\,
        \exp\left[
            -\frac{
                S^{(1)}_{n,\theta}(\lambda')
                -
                S^{(1)}_{n,\theta}(\lambda)
            }{T}
        \right]
        \frac{|R(\lambda)|}{|R(\lambda')|}
    \right\}.
    \label{eq:app_mcmc_acceptance}
\end{equation}
The production calculations use \(T=1\), so the stationary distribution
is the finite-size ensemble defined above.

For each
\(\theta=0,0.1,0.5,\) and \(1\), and each
\(n=500,1000,2000,4000,\) and \(8000\), we run \(64\) independent
chains.  Their initial states are distributed among the MAP candidate,
weakly and strongly jittered MAP candidates, a VKLS-like partition, and
an exponential partition.  Random initial partitions require a
substantially longer burn-in under the adopted local proposal and are
therefore reserved for optional stress tests rather than included in the
production averages.

Each chain is evolved for \(60000\) burn-in moves followed by
\(100000\) production moves.  Every \(250\)th production state is
retained, giving \(400\) samples per chain and \(25600\) retained
partitions for each \((n,\theta)\).

For every retained partition, the English staircase profile is evaluated
on a common cell-centred grid.  The sampled mean profile is
\begin{equation}
    \overline f_{n,\theta}^{\,\rm MCMC}(x)
    =
    \frac{1}{N_{\rm samp}}
    \sum_{a=1}^{N_{\rm samp}}
    f_{\lambda^{(a)},n}(x),
    \label{eq:app_mcmc_mean_profile}
\end{equation}
and the pointwise variance is obtained from
\begin{equation}
    {\rm Var}\,f(x)
    =
    \frac{1}{N_{\rm samp}}
    \sum_{a=1}^{N_{\rm samp}}
    f_{\lambda^{(a)},n}(x)^2
    -
    \left[
        \overline f_{n,\theta}^{\,\rm MCMC}(x)
    \right]^2.
    \label{eq:app_mcmc_profile_variance}
\end{equation}
The same samples are used to measure
\(\lambda_1/\sqrt n\), \(\ell(\lambda)/\sqrt n\), and
\(S^{(1)}_{n,\theta}(\lambda)/n\).

Action, largest-row, and row-number traces are recorded every \(1000\)
production moves.  Chain-to-chain consistency is assessed from the
standard deviations of the corresponding chain-averaged observables.
For brevity, define
\begin{equation}
    \sigma_{\lambda_1}
    =
    {\rm std}_{\rm chain}
    \left(
        \frac{\lambda_1}{\sqrt n}
    \right),
    \qquad
    \sigma_{\ell}
    =
    {\rm std}_{\rm chain}
    \left(
        \frac{\ell(\lambda)}{\sqrt n}
    \right).
    \label{eq:app_mcmc_chain_spreads}
\end{equation}

The diagnostics at the largest sampled size are given in
Table~\ref{tab:app_mcmc_chain_diagnostics}.  The mean acceptance rates
remain between approximately \(0.75\) and \(0.78\), so the chains remain
mobile under the local corner-transfer proposal.

\begin{table}[t]
\caption{
Corner-transfer MCMC diagnostics at \(n=8000\).  The last two columns
give the standard deviations across independent chain means.
}
\label{tab:app_mcmc_chain_diagnostics}
\centering
\begingroup
\renewcommand{\arraystretch}{1.12}
\setlength{\tabcolsep}{8pt}
\begin{ruledtabular}
\begin{tabular}{cccc}
\(\theta\)
&
Mean acceptance
&
\(\sigma_{\lambda_1}\)
&
\(\sigma_{\ell}\)
\\
\hline
\(0.0\) & \(0.7517\) & \(0.0200\) & \(0.0175\) \\
\(0.1\) & \(0.7472\) & \(0.0050\) & \(0.0188\) \\
\(0.5\) & \(0.7650\) & \(0.0026\) & \(0.0225\) \\
\(1.0\) & \(0.7830\) & \(0.0019\) & \(0.0276\)
\end{tabular}
\end{ruledtabular}
\endgroup
\end{table}

The recorded action and shape observables show no systematic drift over
the retained production interval.  Chains initialized from the MAP
candidate, its jittered variants, the VKLS-like partition, and the
exponential partition give consistent mean profiles and closely agreeing
chain-averaged observables.  These diagnostics support the use of the
combined chains in the MCMC--MAP comparisons reported in
Sec.~\ref{subsec:quartic_mcmc}.

\clearpage
\bibliography{Ref}

\end{document}